%% file: main_jrnl.tex
\documentclass[journal, 10pt]{IEEEtran}

\input{structure}
\graphicspath{{figures/}}

\begin{document}

\newcommand{\FP}[1]{{\color{blue}#1}}
\newcommand{\MM}[1]{{\color{red}#1}}

\title{Deinterleaving RADAR emitters with optimal transport distances}

\author{
    Manon Mottier,
    Gilles Chardon,
    Frédéric Pascal~\IEEEmembership{Senior member,~IEEE,}\\
\{manon.mottier, gilles.chardon, frederic.pascal\}@centralesupelec.fr
    \thanks{The authors are with Université Paris-Saclay, CNRS, CentraleSupélec, Laboratoire des signaux et systèmes, 91190, Gif-sur-Yvette, France, and received support for this study from ATOS, which contributed both data and RADAR expertise.}
}

\maketitle

\begin{abstract}
    Detection and identification of emitters provide vital information for defensive strategies in electronic intelligence. Based on a received signal containing pulses from an unknown number of emitters, this paper introduces an unsupervised methodology for deinterleaving RADAR signals based on a combination of clustering algorithms and optimal transport distances. The first step involves separating the pulses with a clustering algorithm under the constraint that the pulses of two different emitters cannot belong to the same cluster. Then, as the emitters exhibit complex behavior and can be represented by several clusters, we propose a hierarchical clustering algorithm based on an optimal transport distance to merge these clusters. A variant is also developed, capable of handling more complex signals. Finally, the proposed methodology is evaluated on simulated data provided through a realistic simulator. Results show that the proposed methods are capable of deinterleaving complex RADAR signals.
\end{abstract}


\IEEEpeerreviewmaketitle





    




\input{./part_introduction.tex}

\input{./part_data_description.tex}

\input{./Deinterlaving_Approach1.tex}
\input{./Deinterlaving_Approach2.tex}

\input{./part_results.tex}

\input{./part_conclusion.tex}

\bibliographystyle{IEEEtran}
\bibliography{bibliography}

\bigskip 

\end{document}

%% file: structure.tex
\usepackage{cite}
\usepackage{amsmath,amssymb,amsfonts}

\usepackage{algorithmic}
\usepackage{algorithm}
\usepackage[algo2e]{algorithm2e} 

\DeclareMathOperator*{\argmin}{argmin}
\DeclareMathOperator*{\argmax}{argmax}
\newcommand{\R}{\mathbf{R}}

\usepackage{graphicx}
\usepackage{textcomp}

\usepackage{xcolor}
\usepackage{colortbl} 

\usepackage{multicol} 
\usepackage{multirow}
\usepackage{subcaption}

\usepackage{enumitem}

\usepackage{hyperref}

\SetKwInput{KwResult}{{Result}}
\SetKwInput{KwData}{{Data}}
\SetKwInput{KwInput}{{Input}}
\SetKwInput{KwOutput}{{Output}}
\SetKwInput{KwProcedure}{{Procedure}}
\SetKwInput{KwParameters}{{Parameters}}
\SetKwInput{KwInitialisation}{{Initialisation}}
\SetKwInput{KwMetrics}{{Metric}}
\SetKwInput{KwFeatures}{{Features}}
\SetKwInput{KwFor}{{For}}

%% file: part_introduction.tex
\section{Introduction}

\IEEEPARstart{I}{}n a passive listening system {\cite{skolnik2008radar, wiley2006elint, wilkinson1985use, davies1982automatic, clarkson1993parallel}}, an Electronic Support Measure system intercepts signals containing interleaved time domain pulses from an unknown number of emitters. Pulse description words (PDW) represent pulses containing various parameters, also called features, such as frequency, pulse width, angle of arrival, etc. Over the years, many methods have been developed to deinterleave a RADAR signal, i.e., to separate the pulses from different emitters. At first, the systems were less sophisticated than now; simple and easily separable characteristics represented emitters as transmitted continuously on a single frequency (little noise, few adverse countermeasures). Early work from the 80s mainly used frequency, direction of arrival, and time of arrival to deinterleave RADAR signals~\cite{wilkinson1985use}. After filtering the frequency and the direction of arrival, the time of arrival is used to find the pulse repetition interval (PRI). The PRI corresponds to the difference in emission between two successive pulses and is usually different between RADAR emitters. This work has been a foundation for many authors who have contributed to enriching the RADAR literature. Most of the following methods are based on the search of patterns in the PRI, for example, by using the cumulative difference histogram (CDIF)~\cite{mardia1989new} or the sequential difference histogram (SDIF)~\cite{milojevic1992improved}.\\

Searching for the PRI pattern is straightforward for simple RADAR emitters because they transmit continuously with a single PRI. Nowadays, emitters transmit less regularly and can have characteristics of agility, such as emitting on several frequencies with one or more PRIs. Some methods have adapted to cope with these changes~\cite {nishiguchi2000improved}. Deep learning models have been increasingly requested to manage the sophistication and emission complexity of emitters~{\cite{liu2018classification, zhou2018automatic, zhu2021model, liu2021pulse, nuhoglu2022image, 9825692, sharma2022pri, 10274853, 10150837, 9844007}}. Deep Learning models require considerable data and are difficult to parameterize. These methods are primarily built in a supervised framework, and their results strongly depend on the quality of the simulator used. Most of these methods are based on processing the PRI data. When the signal is poorly estimated, it is possible to have missing pulses, which can lead to a poor reconstruction of the PRI and, therefore, to failure to regroup the pulses correctly. Indeed, due to the complexity of the problem, supervised approaches, such as deep learning techniques, generally failed to provide good deinterleaving performance. However, methods based on clustering pulse features are based on models too simple to deal with the variety and complexity of RADAR signals~\cite{dong2022distributed, scherreik2021online, 10109668}. Indeed, these methods cannot identify RADAR emitters that emit on several different frequencies, pulse width, etc..\\
 
To tackle more complex cases (\textit{e.g.}, agile emitters, noisy signals, etc.), we propose two simple unsupervised methods based on optimal transport distances to deinterleave a signal without considering the PRI. The first method is based on an algorithm presented at a conference {\cite{9455272} (see also \cite{9909967} for an application of optimal transport to RADAR emitter classification)}, which now integrates cluster processing with a few pulses and a more efficient decision model to cut the dendrogram during the hierarchical clustering. Then, we propose a variant capable of handling emitters with similar characteristics at a low signal-to-noise ratio (SNR). Empirical results are given to evaluate and compare the robustness of the proposed methods under challenging cases (complex emitters characteristics, noisy data, etc.). Note that, for such problems in electronic warfare, data are often provided by MoD services/companies in the defense domain. To cope with such restrictions, this work provides a data simulator that mimics real data.\\

The paper is structured as follows: the realistic simulated dataset is presented in Section \ref{sec:data-description}. Section \ref{ssec:AP1step0} introduces the proposed methodology with the basic notion of optimal transport, followed by an improved version of this algorithm in Section \ref{ssec:AP2step0}. Simulation results are given in Section \ref{ssec:results} before concluding.

%% file: part_data_description.tex
\section{Data description}\label{sec:data-description}

This section introduces the structure of the data considered in this work. Due to the complexity of working with real and labeled data (for defense security matters), the proposed deinterleaving methods are tuned, parameterized, and validated on simulated data. The complexity of the simulated data represents the main challenge in RADAR signal deinterleaving. We have created a RADAR signal simulator based on signal theory and telecommunication engineering. The RADAR simulator assumes the receiver is static, has a known detection threshold, and is omnidirectional, making the direction of arrival unusable. Several characteristics (or features) describe each emitter class:

\begin{itemize}
    \item Pulses parameters: frequency, pulse width, level, direction of arrival, and pulse repetition interval pattern.
    \item Technical emitter parameters: transmitted power,  scan period, antenna directivity, and spatial distance from the receptor.\\
\end{itemize}
 As the received signal is noisy, the features
 are perturbed by estimation errors. In addition, some low-energy pulses may be lost or truncated.

There is a wide variety of emitters. Some may have elementary parameters, such as emitting on a single frequency, while others have sophisticated pulse characteristics, \textit{e.g.}, agility in frequency, multiple pulse repetition of intervals, etc., or with rarer appearance periods. From these heterogeneous emitter profiles, an interleaved signal is created. The simulator controls the number of emitters in the signal and the desired number of pulses. A listening system collects the data without information about the environment. The diversity of the simulated signals results in the acquisition of single- or multi-sensor labeled signals, signals with frequency or time modulation, and signals comprising measurement errors, outliers, missing data, or non-Gaussian noise. The Doppler effect, fundamental in classical RADAR processing, is neglected here, as the frequency shifts corresponding to realistic target velocities are negligible compared to the difference between frequencies used by different emitters. {Another source of uncertainty in the measurements, multipath propagation, manifests itself by additional pulses, sharing the same frequency and duration as the original pulses, with decreased power and slightly shifted in time. As the proposed methods are based on the
unaffected frequency and duration of the pulses, multipath propagation will be neglected.}



\begin{figure}[htbp]
\centerline{\includegraphics[width = 8.5cm]{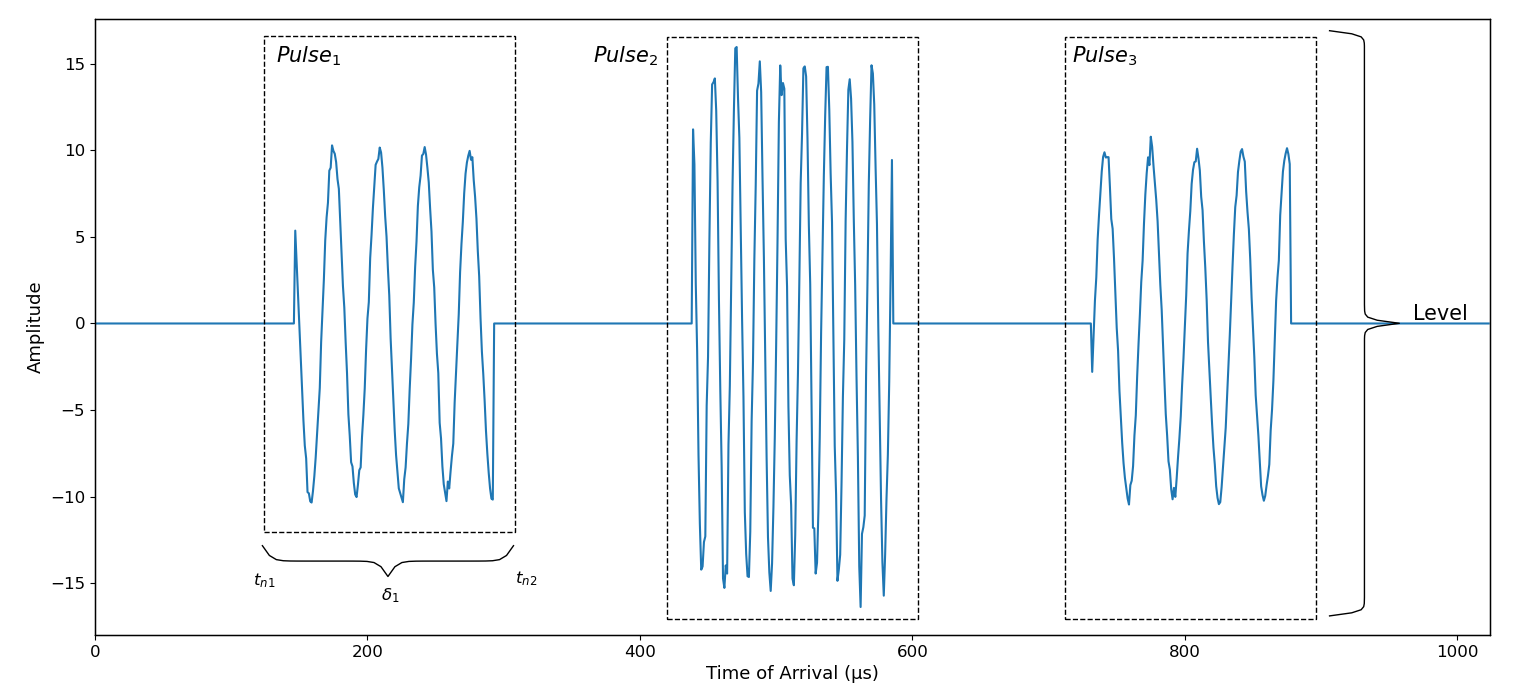}}
\caption{Example of three simulated pulses from an emitter.}
\label{fig_DS:pdw_brut}
\end{figure}

A short simulated signal is shown in Fig.~\ref{fig_DS:pdw_brut}, containing three pulses. After the signal acquisition, the pulses are segmented, analyzed, and described by features. In this work, the $n$-th acquired pulse is described using the following four characteristics:

\begin{itemize}
    \item Frequency, F ($f_n$)
    \item Pulse width, PW ($pw_n$)
    \item Level, G ($g_n$)
    \item Time of Arrival, TOA ($t_n$)\\
\end{itemize}

\vspace{-0.2cm}

An additional feature can be added: the difference of time of arrival (DTOA) $\delta_n$, defined as the interval of time between two successive pulses belonging to the same emitter (in the case of a unique emitter, $\delta_n = t_n-t_{n-1}$). As the emitters are not separated at this stage, two successive pulses are not guaranteed to be transmitted by the same emitter. As a consequence, DTOA cannot be used before deinterleaving the signal.\\

\begin{figure}[h]
\centerline{\includegraphics[width = 9cm]{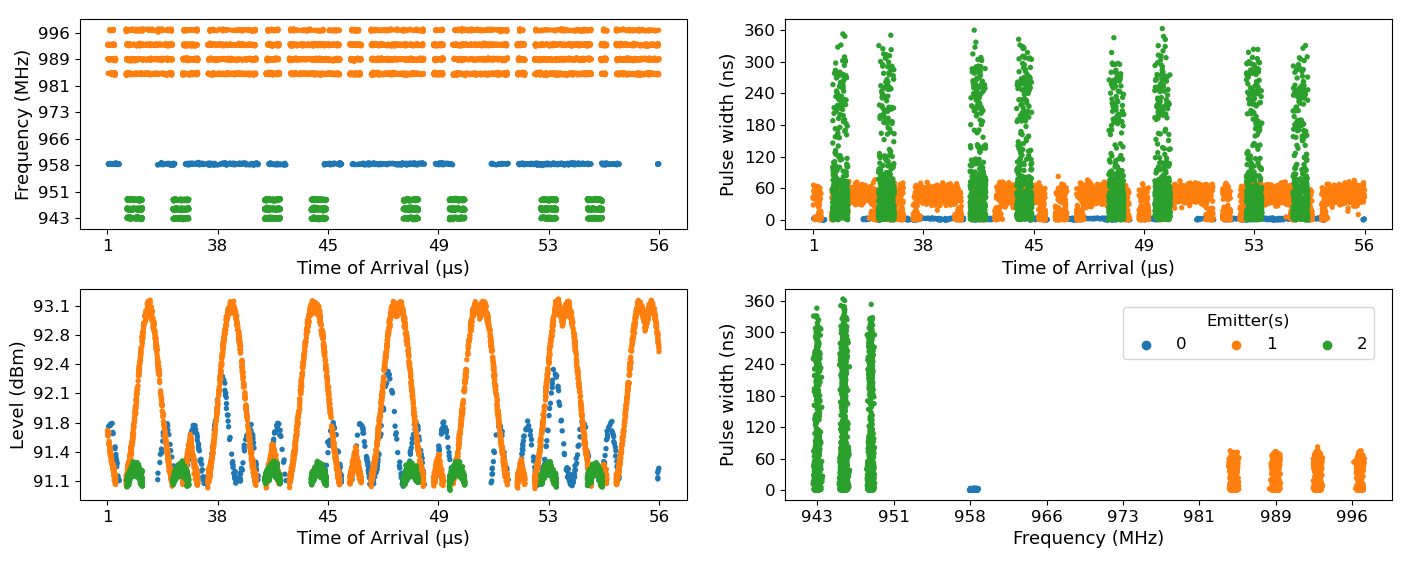}}
\caption{Example of a simulated signal gathering 8917 pulses of three emitters, identified by a color.}
\label{fig_DS:data_true_label}
\end{figure}

Other features (waveform, frequency modulation, etc.) will not be considered here. Our approach considers only the four above-mentioned features because they are always available and relatively well-estimated. Fig~\ref{fig_DS:data_true_label} shows a simulated signal containing 8917 pulses of three different emitters, each represented by a color (blue, green, and orange). Each point represents a pulse. Frequency, level, and pulse width are plotted in function of time in panels Fig.~\ref{fig_DS:data_true_label}-a, -b, and -c, respectively. Additionally, in Fig.~\ref{fig_DS:data_true_label}-d, pulses are represented in the $(f_n, pw_n)$ plane. These representations highlight several challenging characteristics of RADAR signals:


\begin{itemize}
    \item Several emitters can be active at the same time.
    \item Emitters can transmit continuously and regularly on different frequencies, as shown by emitter 1 in the $(f_n, t_n)$ plane, which transmits on four different frequencies or emitter 2 on three frequencies.
    \item Pulse widths can be severely misestimated, as low-power pulses can be split during the segmentation stage, spreading the pulses over 360 ns.
    \item In general, estimated features are noisy.\\
\end{itemize} 

In particular, several clusters in the $(f_n, pw_n)$ plane can represent a given emitter, necessitating appropriate processing to regroup these clusters specifically for clustering-based methods in this plane. Our proposal introduces methods based on optimal transport distances capable of capturing and managing this phenomenon. 

%% file: Deinterlaving_Approach1.tex
\section{Hierarchical agglomerative clustering combined with optimal transport distances}\label{ssec:AP1step0}

As pointed out in the previous section, one of the challenges in deinterleaving is the dispersion through several groups of pulses from an emitter in the $(f_n, pw_n)$ plane. {To overcome this problem, an extension of a two-step strategy proposed by the authors in~\cite{9455272} is introduced and highlighted in Algorithm~\ref{algo_AP1:hacot}}. Firstly, a clustering algorithm separates the pulses in the $(f_n, pw_n)$ plane. Then, a hierarchical agglomerative clustering algorithm combined with optimal transport distances groups these clusters as an emitter can be represented by several clusters.\\

\begin{algorithm}[h]

    \caption{Hierarchical agglomerative clustering using optimal transport distances to deinterleave emitter pulses - HACOT}
    \label{algo_AP1:hacot} \medskip
    
    \KwData{Set of pulses ($X$)} \medskip

    \KwFeatures{Frequency ($f_n$), Pulse width ($pw_n$), Level ($g_n$), Time of arrival ($t_n$)} \medskip
    
    \KwParameters{Minimum number of points to form a cluster ($\textrm{minPts}$), Confidence level ($\alpha$), Statistical test ($test$), Non-parametric method ($method$), Threshold ($\lambda$))} \medskip

     \KwProcedure{     
     \begin{enumerate}
        \item Pulses separation: apply HDBSCAN for all $x_i$ of $X$ from $f_n$ and $pw_n$ according to $\textrm{minPts}$: $C$ (Set of clusters) \medskip

        \item Cluster aggregation: apply hierarchical agglomerative clustering based on optimal transport distances presented in algorithm~\ref{algo_AP1:cluster_aggregation} considering $\alpha$, $test$, $method$ and $\lambda$ for all $c_i$ of $C$ from $t_n$ and $g_n$: $Y$ (Set of aggregated clusters) 
        \medskip      
        
    \end{enumerate}}
    
    \KwResult{$Y$, Deinterleaved sets of pulses.} \medskip
       
\end{algorithm}


\subsection{Pulses separation with HDBSCAN in 2 dimensions}\label{ssec:AP1step1}

The first step of the method is to apply a clustering algorithm to separate the pulses only from two highly discriminating and reliable characteristics: frequency and pulse width. It is important to note that not merging two different emitters in the same cluster is crucial. Thus, the proposed approach will overestimate the number of clusters at this step. {Several algorithms were tested to determine the most suitable for the method, such as K-MEANS~\cite{hartigan1979algorithm}, Hierarchical Agglomerative Clustering (HAC)~\cite{johnson1967hierarchical}, Gaussian Mixture Models (GMM)~\cite{mclachlan1988mixture}, Bayesian Gaussian Mixtures Models (BGMM)~\cite{roberts1998bayesian}, or density-based algorithms (DBSCAN~\cite{ester1996density}, OPTICS~\cite{ankerst1999optics}, HDBSCAN\cite{campello2013density}). K-MEANS and GMM require assumptions about the number of clusters to be identified. Conventional methods such as the Elbow method, unsupervised metrics, or data projection can set this number. These methods are effective in handling uncomplicated signals and emitters with simple characteristics. However, deinterleaving algorithms are designed for electronic warfare and unsupervised frameworks, where emitters often have complex and similar characteristics, thus rendering these methods unsuitable. Manually fixing the number of clusters for detection is unfeasible because each signal differs.} 

We propose to use the Hierarchical Density-Based Spatial Clustering of Applications with Noise algorithm (HDBSCAN)\cite{campello2013density}, an unsupervised clustering algorithm based on a hierarchical version of the Density-based spatial clustering of applications with Noise (DBSCAN)\cite{ester1996density}. {The choice of the unsupervised HDBSCAN algorithm is motivated by its capacity to identify clusters with different densities and shapes, the few parameters to optimize, its ability to detect the clusters without setting the number of clusters to be detected beforehand, and its capacity to clean the signal by detecting outliers.} 

{The main idea is to group points that live in the same dense area from a vector $ X = \{(f_1, pw_1), (f_2, pw_2), ..., (f_N, pw_N)\}$ grouping frequency and pulse width pulses from a signal of $N$ pulses.} Clustering results strongly depend on the hyperparameter determining the minimum size of a group that can be considered a cluster. HDBSCAN has been configured to overestimate the number of clusters returned so as not to mix the pulses of several emitters in the same cluster. Comparisons were made to find a tradeoff to fix this threshold. HDBSCAN is applied to normalized data using the quantile method in the $(f_n, pw_n)$ plane because it preserves the data distribution as well as possible. \\





\begin{figure}[htbp]
\centerline{\includegraphics[width = 8cm]{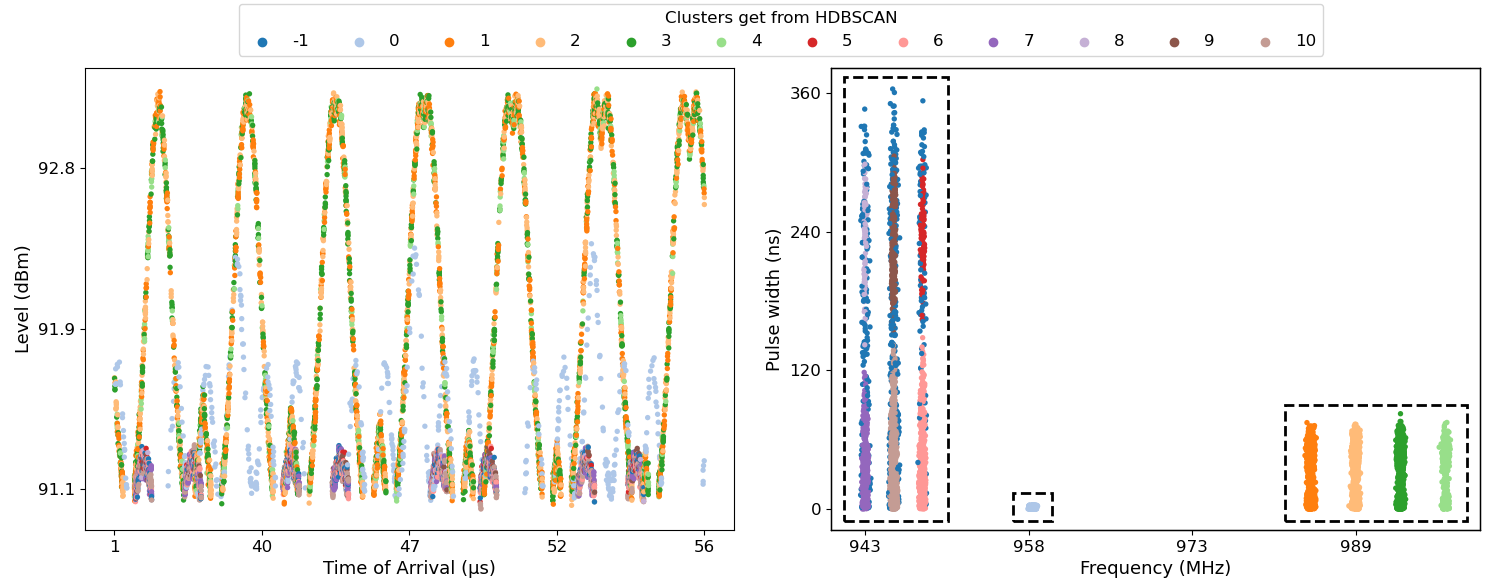}}
\caption{HDBSCAN outputs obtained from frequency and pulse width. The algorithm identifies 11 clusters and an outliers class (-1), and the boxes highlight the pulses of the three emitters.}
\label{fig_AP1:DG1_data_clustering}
\end{figure}

The results of the clustering performed by HDBSCAN in the $(f_{n}, pw_{n})$ plane are shown for the simulated dataset presented in Section~\ref{sec:data-description} in Fig.~\ref{fig_AP1:DG1_data_clustering}. The algorithm identifies 11 clusters and a class of outliers (-1) with heterogeneous sizes ranging from 86 to more than 1500 pulses. The $(f_{n}, pw_{n})$ plane shows the plane where the clustering was performed and that the clustering results are consistent; the black boxes highlight the present emitters. The pulses from emitter 0 transmitting around 958 MHz have been correctly grouped into a single cluster. Conversely, HDBSCAN splits the emitter pulses above 981 MHz and below 951 MHz into several clusters. Indeed, both emitters transmit on several frequencies. Additionally, low-power pulses of the lower frequency emitter are split at the segmentation stage, adding short pulse widths in the data.

The $(g_{n}, t_{n})$ plane on the left highlights the distribution of these clusters along the lobes.

\subsection{Cluster fusion with hierarchical agglomerative clustering using optimal transport distances}\label{ssec:AP1step2}

As shown in Fig.~\ref{fig_AP1:DG1_data_clustering}, an emitter can be split over several clusters. Therefore, the second step of the method is designed to group these clusters according to shared characteristics. In particular, we assume that the clusters of a given emitter are simultaneously active, as illustrated in Fig.~\ref{fig_AP1:DG1_data_clustering} where the pulses of an emitter are distributed over the lobes. From the HDBSCAN clusters, hierarchical agglomerative clustering was performed\cite{chakraborty2020hierarchical} using the optimal transport distances \cite{villani2009optimal,bonneel2011displacement} from time of arrival and level as illustrated in  Algorithm~\ref{algo_AP1:cluster_aggregation}. The rest of the article will show that optimal transport distances are particularly well-suited to our problem.\\



\begin{algorithm}[h]

    \caption{Cluster aggregation with hierarchical agglomerative clustering based on optimal transport distances}
    \label{algo_AP1:cluster_aggregation} \medskip
    
    \KwData{Set of clusters ($C$)} \medskip

    \KwFeatures{ Level ($g_n$), Time of arrival ($t_n$)} \medskip

    \KwParameters{Threshold ($\lambda$), Non-parametric method ($method$), Confidence level ($\alpha$), Statistical test ($test$)} \medskip 

    \KwProcedure{\\

     \begin{enumerate}
     
        \item Cluster significance analysis: separation of clusters according to $\lambda$: $D = \{c_i \in C \mid |c_i| \geq \lambda \}$ and $E = C \setminus D$ \medskip

        \item Determination of a hierarchical structure aggregating the clusters using optimal transport distances from $D$: $structure$\\

        \textbf{While} $len(D) \neq 1$ \textbf{do}
        \begin{itemize} \item[]
        \begin{itemize}
            \item For all $d_i$ of $D$, representation by a measure from \\ $t_n$ weighted by $g_n$: $\tau_{i}$
            
            \item Computation of the distance for all $d_i$ of $D$ using the optimal transport: $d\left(\tau_{i}, \tau_{j}\right)$\\
            
            \item Aggregation of the two closest clusters: $(i, j)^\star$\\
            \item Updating $D$  
            
        \end{itemize}
        \end{itemize}
        \textbf{end}
        
        \medskip
        
        \item Dendrogram pruning according to the decisional model presented in Algorithm~\ref{algo_AP1_algo:DecisonalModel_test} considering $method$ and $test$: set of aggregated clusters, $F$ \medskip

        \item Dealing with excluded clusters $E$: 

        \begin{itemize}
            \item Estimation of the probability density for all $f_i$ in $F$ from $t_n$ using $method$

            \item Association of all $e_i$ of $E$ to $F$ by maximum likelihood estimation

            \end{itemize}
        
    \end{enumerate}

    }

    \KwOutput{$F$, aggregated clusters.} \medskip
     
\end{algorithm}  

To define optimal transport distances between clusters, each cluster is represented by a probability measure describing its distribution from time of arrival weighted by level:

\begin{equation}
    \tau = \frac{1}{Z} \sum_{p = 1}^P g_{p} \delta_{t_{p}}.    \label{eq_ot:measure_deinterlaving_cluster}
\end{equation} 

with $P$, the number of pulses in the cluster and $Z = \sum_{p = 1}^P g_{p}$, the total energy in the cluster.
{In order to decrease the computational complexity of the method and avoid numerical problems, the probability measures used in practice are obtained from data histograms. The range of time of arrivals is partitioned in $B$ intervals $[b_{i-1}, b_i)$. The measure $\tau$ in (\ref{eq_ot:measure_deinterlaving_cluster}) is replaced by
\begin{equation}
    \bar{\tau} = \sum_{i=1}^B \bar{g}_i \delta_{c_i}. \label{eq_ot:measure_deinterlaving_cluster_histo}
\end{equation} 

where $c_i = \frac{1}{2}(b_{i-1} + b_i)$, and $\bar g_i$ is the energy of the pulses in the interval, normalized so that $\sum_{i=1}^B \bar g_i =1$.


The number of the bins is fixed according to the Freedman–Diaconis rule~\cite{freedman1981histogram}: $B = 2\frac{\mathrm{IQ}(N)}{\sqrt[3]{N}}$ with $N$ the signal size and $\mathrm{IQ}$ the interquartile range of the signal. This method is less sensitive to outliers because it is based on the width of the IQ range, which is less influenced by outliers than the mean and standard deviation. In addition, it better adapts to non-normal or skewed distributions by considering the real dispersion of the signal. This method adjusts for the sample size, avoiding over or under-determination of the number of bins (other methods could be considered). Finally, the bin values are defined uniformly concerning the time of arrival values of the signal.}\\

{We now recall the definition of optimal transport distances, or Wasserstein-Kantorovitch distances, in the particular case of discrete probability measures.} Let us consider two discrete probability distributions: $$\nu = \sum_{n=1}^N a_n \delta_{x_n} \text{ and } \mu = \sum_{m=1}^M b_m \delta_{y_m},$$ with $\mathbf a = (a_1, \ldots a_N)^T \in \R_+^N$, such that $\sum_{n=1}^N a_n = 1$, and $\mathbf b = (b_1, \ldots b_M )^T \in \R_+^M$, such that $\sum_{m=1}^M b_m = 1$.\\


A transport plan $\mathbf P$ between $\nu$ and $\mu$ is defined by its coefficients $P_{nm}$, representing the amount of mass taken from $x_n$ to $y_m$. Consider the cost function $c(\cdot, \cdot)$. In this work, we use the $L_2$-norm: $c(x, y) = \| x - y\|_{2}$. 

$C_{nm} = c( x_n, y_m)$ represents the cost of transporting a unit of mass from $x_n$ to $y_n$, and the total cost $C(\mathbf P)$ of a transport plan is:\\

\begin{equation}
        C(\mathbf P) = \sum_{n=1}^N \sum_{m=1}^M C_{nm} P_{nm}. 
        \label{eq_ot:transport_plane}
\end{equation}

The consistency of the transport plan $\mathbf P$ with $\nu$ and $\mu$ is guaranteed by $\mathbf P \mathbf 1_M = \mathbf a, \mathbf P^T \mathbf 1_N = \mathbf b$. The optimal transport plan $\mathbf P^\star$ is defined as the minimizer of Eq.~(\ref{eq_ot:transport_plane}) under the following constraints:
\begin{equation}
    \mathbf P^\star = \argmin_{\mathbf P \in \R_+^{N\times M}} C(\mathbf P)
    \mbox{ subject to } \mathbf P \mathbf 1_M = \mathbf a, \mathbf P^T \mathbf 1_N = \mathbf b.
    \label{eq_ot:transport minimization}
\end{equation}

\begin{figure}[h]
\centerline{\includegraphics[width = 9cm]{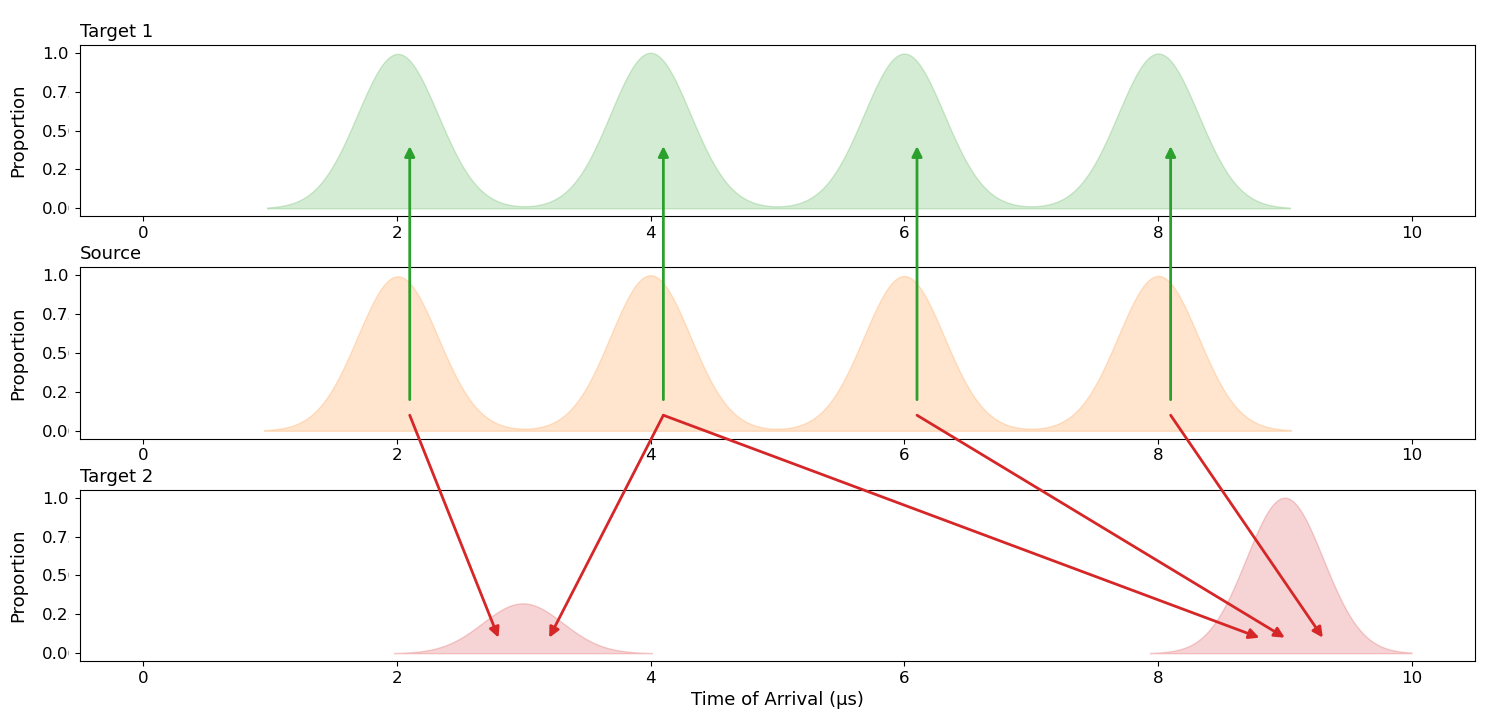}}
\caption{Example of three simulated TOAs distribution. Source and Target 1 belong to the same emitter and are characterized by a low transport cost. Conversely, Source and Target 2 come from 2 different emitters, implying a high transportation cost due to the difference in transmission times.}
\label{fig_DS:ot_explanation_deinterlaving}
\end{figure}

The optimal transport distance between $\nu$ and $\mu$ is then defined by $d(\nu, \mu) = C(\mathbf P^\star)$. Fig.~\ref{fig_DS:ot_explanation_deinterlaving} displays an example of applying optimal transport to define a distance between three simulated TOA distributions. Source and Target 1 belong to the same emitter and are active simultaneously, requiring very few moves to make them match, so low transport costs characterize them. Conversely, Source and Target 2 come from 2 different emitters, implying a high transportation cost due to the difference in transmission times.

{In this particular case of univariate distributions, the distance can be computed using the cumulative distribution functions or quantile functions of the distributions. Note that the formulation (\ref{eq_ot:transport minimization}) is more general, as it can be extended to multivariate distributions and can be modified to deal with unbalanced distributions \cite{SEJOURNE2023407}.}

The optimal transport distance is used in an agglomerative clustering algorithm; the two clusters with the smallest optimal transport distance are aggregated:

\begin{equation}
    (i^\star, j^\star) = \argmin d\left(\tau_{i},\tau_{j}\right).
    \label{eq_ot:classification_decision}
\end{equation} 

After merging, the distances between merged clusters and other clusters are updated. Then, the process is repeated until the clusters are fully aggregated. To obtain accurate estimates of the distance between clusters, each cluster must have sufficient pulses, as a probability distribution represents them. {This threshold was set to 100, considering the results we have seen on several signals.} Clusters having fewer pulses than this threshold are excluded from the hierarchical approach but will be processed separately using an alternative approach afterward. Here, cluster 8, with 86 pulses, is set aside.\\

\begin{figure}[h]
\centerline{\includegraphics[width = 9cm]{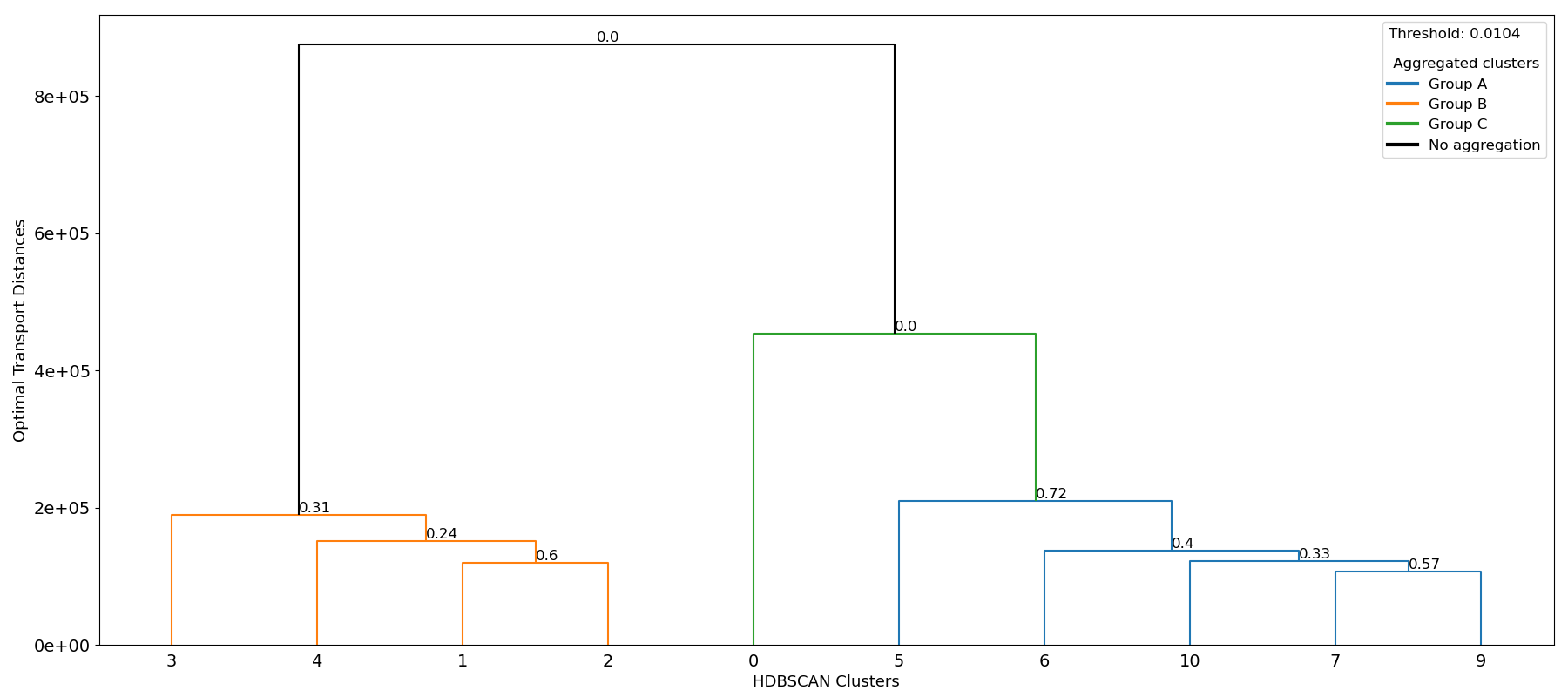}}
\caption{Dendrogram representing the aggregations of hierarchical agglomerative clustering combined with optimal transport distances. The $p$-values of the Kolmogorov-Smirnov test from $t_n$ evaluating the aggregated groups are displayed at each aggregation, while each color identified the final group.}
\label{fig_AP1:dendrogram}
\end{figure}

Fig.~\ref{fig_AP1:dendrogram} represents the aggregation at each step of the hierarchical agglomerative clustering combined with the optimal transport distances, with heights of the edges representing the optimal transport distances calculated at each aggregation. The final clustering is obtained by stopping the fusion of clusters at appropriate stages. Indeed, prematurely stopping the fusions may result in separating several emitters' pulses into multiple groups, and late fusion results in grouping pulses of different emitters in a single cluster.

\begin{figure}[h]
\centerline{\includegraphics[width = 9cm]{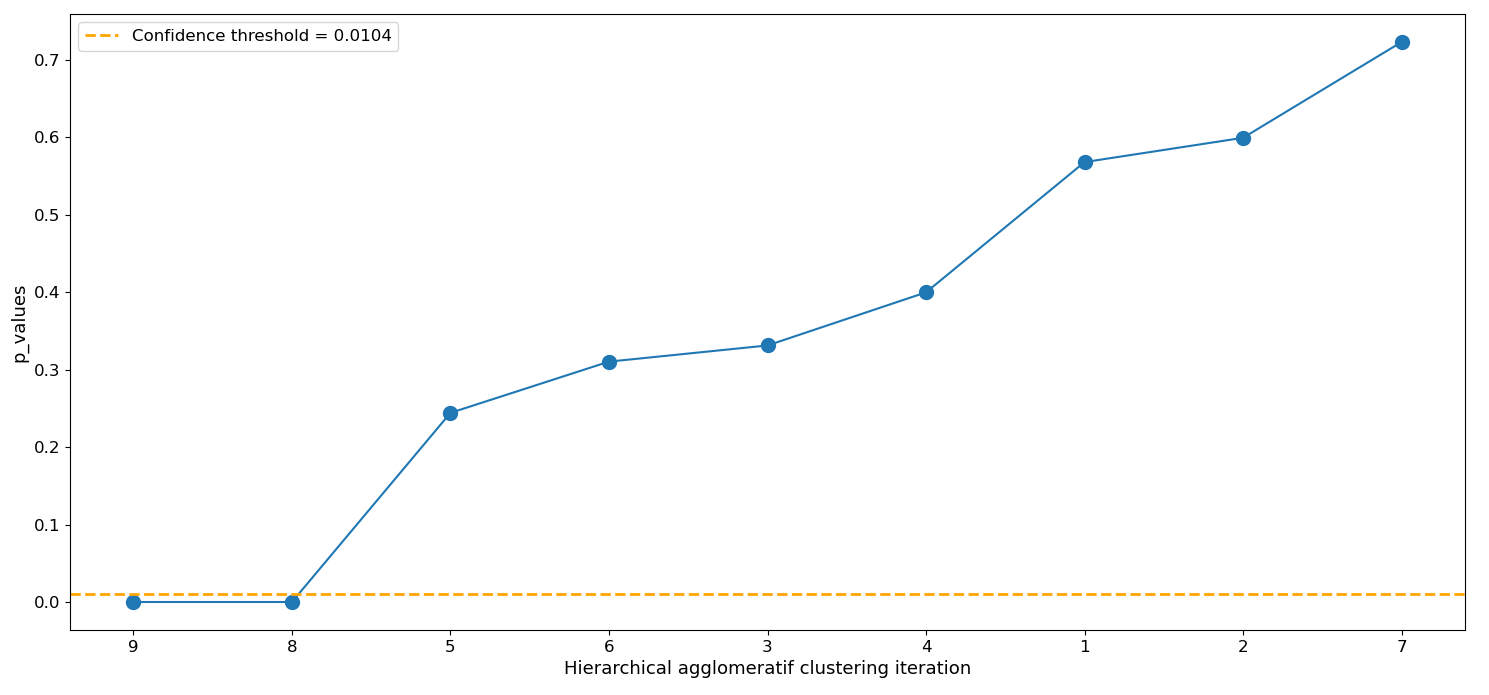}}
\caption{$p$-values sorted according to the Kolmogorov-Smirnov test, and the orange line highlights the estimated confidence level.}
\label{fig_AP1:elbow_trics}
\end{figure}

{This final clustering is obtained by considering each fusion by increasing distance and deciding if this fusion should actually be performed by testing if the involved TOAs are likely to be sampled from the same TOA probability distribution. If the fusion is not performed, the later fusions involving these clusters are not considered. We propose to use the Kolmogorov-Smirnov test~\cite{massey1951kolmogorov} to decide if a fusion is to be performed. For each fusion, the $p$-value of the hypothesis that the two clusters have been drawn from the same distribution is computed. Fusions with a $p$-value higher than a confidence threshold are performed. 

This threshold is estimated by sorting the $p$-values in increasing order and identifying a breakpoint. The $p$-values for the test case considered here are plotted in figure \ref{fig_AP1:elbow_trics}, and the threshold is fixed at $\alpha_{ks} = 0.0104$, indicated by the horizontal line.}

 Given the dendrogram in Fig.~\ref{fig_AP1:dendrogram}, Clusters 7 and 9 are first evaluated and are represented by a measure describing their distributions as in Eq. ~\eqref{eq_ot:measure_deinterlaving_cluster}. The associated $p$-value is 0.57, which is higher than the threshold, meaning clusters 7 and 9 belong to the same emitter and are aggregated. The process is repeated until the top of the tree is reached, or a non-significant $p$-value is returned. The $p$-value between clusters 0 and 5, 6, 10, 7, and 9 is 0, which means the clusters should not belong to the same emitter. The process stops, and the next iteration is evaluated. The decisional model identifies three aggregated groups.\\

Compared to the method proposed in \cite{9455272}, where fusions were stopped
at a given transport optimal distance, obtained by considering three unsupervised metrics, the fusions are here analyzed locally. This allows us to handle more complicated cases, e.g., with heterogeneous emitters, where a fusion should be performed, even if the associated optimal transport distance is higher than a fusion for which fusion was not performed.

\begin{algorithm}[h]
    \caption{Improved decisional model for pruning the dendrogram based on statistical test.}
    \label{algo_AP1_algo:DecisonalModel_test} 
    
    \KwData{Set of clusters ($D$), Hierarchical structure ($structure$)} \medskip

    \KwFeatures{Time of arrival ($t_n$)} \medskip
    
    \KwParameters{Statistical test ($test$), Confidence level ($\alpha$)} \medskip

    \KwInitialisation{Create a list with all aggregations labels from $structure$ and an empty list: $N$ and $L$} \medskip

    \KwProcedure{\\

    \While{$len(N) \neq 0$}{

    Selection of the first two aggregated clusters according to $structure$: $\{i, j\}$ in $N$\\
    
    Representation by a measure from $t_n$: $\theta_{i}$ and $\theta_{j}$\\
    
    Computation of $test$ between $\theta_i$ and $\theta_j$ and get $p$-value\\

    \eIf{$p$-value $< \alpha$}{
           Remove all aggregations from $N$ using these clusters\\
           }{
           Add $\{i, j\}$ in $L$ and del $\{i, j\}$ from $N$ \\
          }}
          
    Aggregate $l_i$ in $L$ according $structure$: $F$ \medskip
    }

    \KwOutput{$F$: Set of aggregated clusters} \medskip
     
\end{algorithm}



\begin{figure}[h]
\centerline{\includegraphics[width = 9cm]{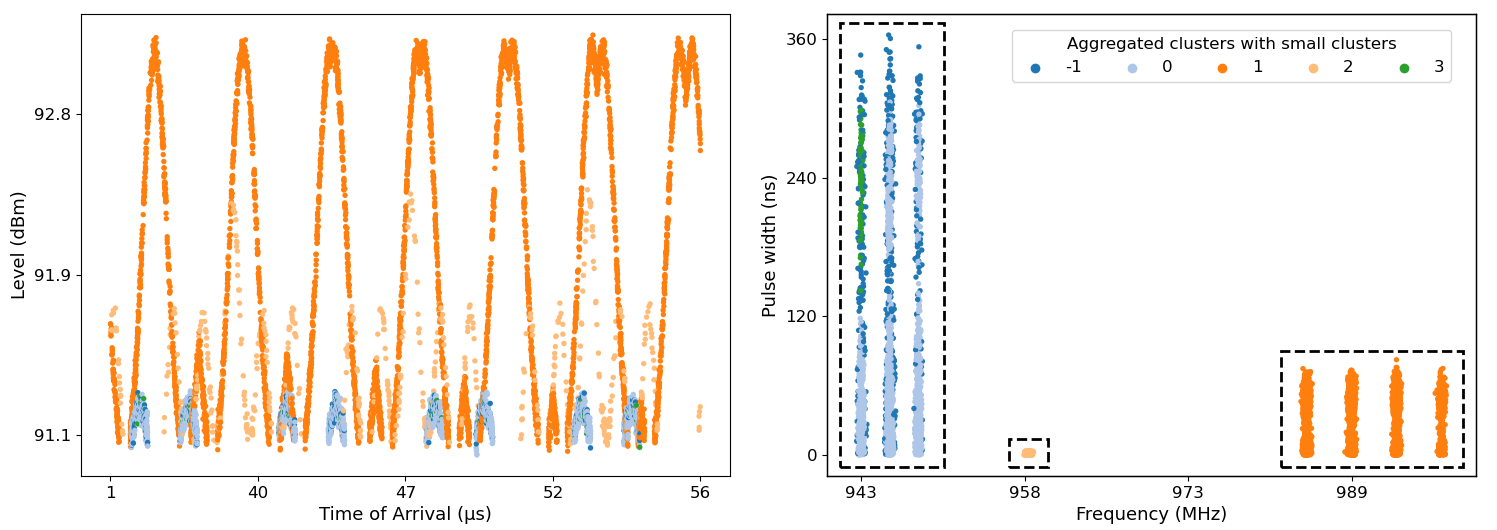}}
\caption{Aggregated cluster obtained following the dendrogram from the hierarchical agglomerative clustering using optimal transport distances. Each color identified an aggregated cluster with an outliers class (-1).}
\label{fig_AP1:data_aggregate_grouping}
\end{figure}

Finally, Fig.~\ref{fig_AP1:data_aggregate_grouping} shows the result of cluster agglomeration. The emitter pulses around 958 Mhz and above 981 Mhz have been perfectly grouped into two sets of pulses. Conversely, the lower frequency emitter is still split into two clusters, as one of the clusters has been excluded from the processing. 

\subsection{Dealing with excluded clusters}\label{ssec:AP1step3}

As previously mentioned, cluster 3 was set aside from the analysis due to its insufficient number of pulses. Clusters excluded from hierarchical clustering are now associated with an extensive cluster obtained after the grouping phase. The process is done in two steps:

\begin{enumerate}
    \item Estimation of the probability density of time of arrival of the extensive clusters using a kernel density estimator~\cite{rosenblatt1956remarks, parzen1962estimation} with Gaussian kernel.
    
    \item Association of an excluded cluster to an extensive cluster by maximum likelihood estimation.\\
\end{enumerate}

We used a non-parametric model to estimate the distribution parameters and their density function. Among these methods, the most commonly used are that of histograms\cite{scott1979optimal, parzen1962estimation}, K-nearest-neighbors\cite{mack1979multivariate,fukunaga1973optimization}, Kernel Density Estimation\cite{scott1980nonparametric} or Neural Networks\cite{magdon1998neural}. We chose to use the kernel density estimation (KDE) because it allows us to build a density from each point and, therefore, to better consider the distribution's behavior. For a sample of points $(x_0, x_1, ..., x_N)$ belonging to a cluster distribution $f$, the kernel density estimator is written as:
\begin{alignat}{1}
   \hat{f}(x) = \frac{1}{Nh} \sum _{n=1}^{N}K\left(\frac{x - x_n}{h}\right), 
\label{eq_AP1:kde}
\end{alignat}
with $N$ the number of pulses, $K$ the kernel, and $h$ the smoothing parameter (bandwidth). The KDE mainly depends on two parameters: the choice of the kernel and the bandwidth. For a Gaussian kernel, by assuming the time of arrival can be represented by a Gaussian distribution, the optimal bandwidth can be obtained from the Scott method's \cite{scott2015multivariate} and is given by:
\begin{equation}
\begin{aligned}
    h &= \left( \frac{4}{3n} \right)^{1/5}{\hat {\sigma}} \\
     &\approx 1.06 {\hat \sigma} n^{-1/5},
\label{eq_AP1:kde_bandwidth}
\end{aligned}
\end{equation}
with $\hat\sigma$ the standard deviation of the cluster. 
As the distribution of the TOA for a given emitter is multimodal, we assume that the distribution of the TOA is a mixture of Gaussian densities.
Temporal clustering with HDBSCAN for each extensive cluster is applied to identify each appearance and estimate the variance of each sub-cluster. The parameter $\hat\sigma$ in the bandwidth estimation is then obtained by averaging the variance of each sub-cluster. Finally, for a given excluded cluster, the likelihood of belonging to an extensive cluster $i$ is given by:
\begin{equation}
    L_i = \prod_{i = 1}^{N} {\hat f_{i}}(t_n),
    \label{eq_AP1:kde_likehood}
\end{equation} 
with $t_n$ the time of arrival. The excluded cluster is assigned to the extensive cluster, maximizing the likelihood:
 \begin{equation}
    i^\star = \argmax L_i,
    \label{eq_AP1:kde_decision}
\end{equation} 

With the likelihood of the considered cluster belonging to the extensive cluster $i$. Cluster 3 will finally be associated with Cluster 2.

\begin{figure}[h]
\centerline{\includegraphics[width = 9cm]{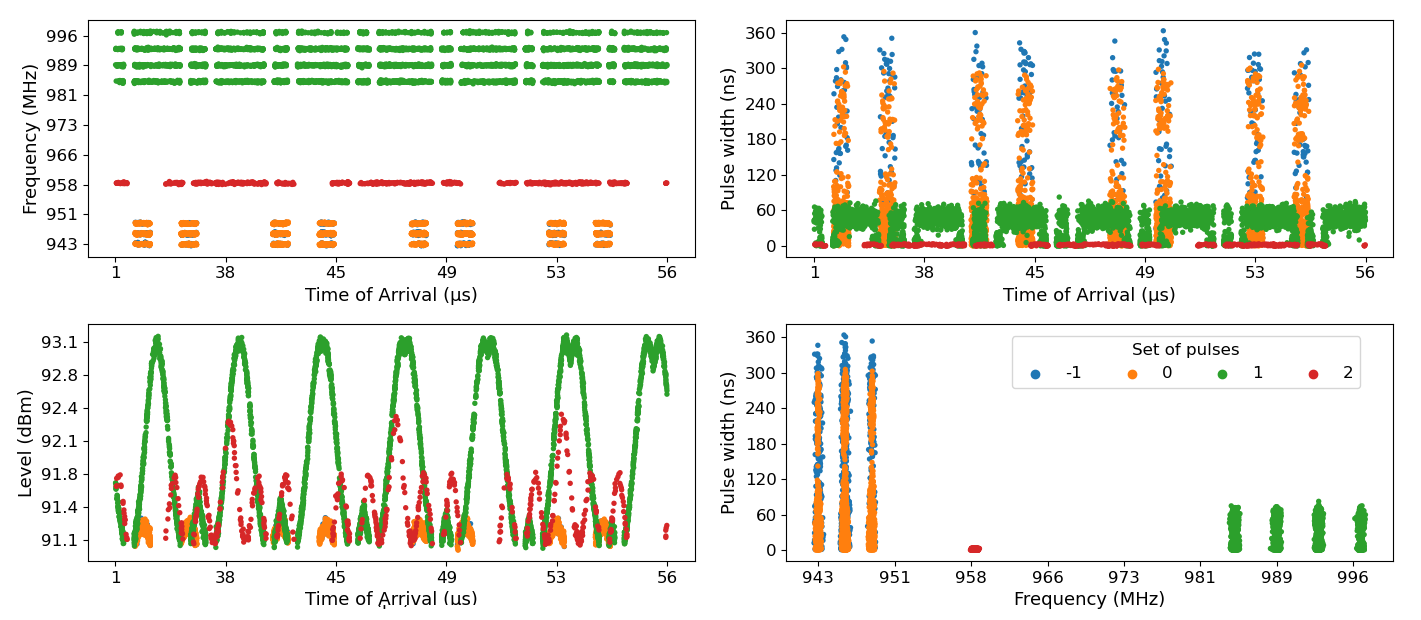}}
\caption{Final grouping after applied hierarchical agglomerative clustering combined with optimal transport distances and kernel density estimation on excluded clusters. Each color identified a set of pulses with an outliers class (-1).}
\label{fig_AP1:data_final_grouping}
\end{figure}

Finally, Fig.~\ref{fig_AP1:data_final_grouping} displays the result of cluster agglomeration. The general hierarchical agglomerative clustering combined with the optimal transport distances and the excluded clusters processing correctly separates the pulses in the signal, groups them into three sets of pulses, and can handle clusters with few pulses. The outlier rate indicates that 4.5\% of the pulses are lost during the deinterleaving process; HDBSCAN excluded these pulses in step 1.


    
    



\subsection{{Computational complexity}}\label{ssec:AP1step4}

{The computational complexity of the first method can be estimated in function of the number $N$ of pulses, the number of clusters $C$ identified by HDBSCAN, and the number of bins $B$. The complexity of the first stage, dominated by HDBSCAN, is $O(N^2)$ \cite{10.1145/2733381}. In the second stage, $C^2$ distances are computed at the beginning of the hierarchical clustering, each costing $O(B)$ using the formulation of the Wasserstein distance as a distance between quantile functions. The number of updated distances during the clustering is also in $O(C^2)$.

The actual clustering, with complexity in $O(C^3)$, can be neglected under the reasonable assumption that the number of bins $B$ is larger than the number of clusters. The $C-1$ Komolgorov-Smirnov tests and the treatment of the excluded clusters are negligible compared to the initial distance computation.

The total complexity of the algorithm is thus $O(N^2 + C^2B)$. This complexity can be rewritten in functions of parameters related to the characteristics of the emitters. With $T$ the length of the signal, $K$ the number of emitters, $\bar R$ the mean rate of impulses, $\bar C$ the mean number of clusters per emitter (i.e., an evaluation of the complexity of the emitter), the complexity is $O(K^2(T^2 \bar R^2 + \bar C^2B))$.}

%% file: Deinterlaving_Approach2.tex
\section{Improved hierarchical agglomerative clustering combined with optimal transport distances}\label{ssec:AP2step0}

In some cases, pulses associated with two different emitters can be grouped in the same cluster by HDBSCAN. This confusion can be caused by the emitters having similar characteristics and inaccurate estimation of the parameters of the pulses. This section proposes a variant of the method to alleviate this issue. Detailed in Algorithm \ref{algo_AP2:ihacot}, it differs from the previous method by adding the time of arrival in the first clustering steps.\\


\begin{algorithm}[h]

    \caption{Improved hierarchical agglomerative clustering using optimal transport distances to deinterleave emitter pulses - IHACOT}
    \label{algo_AP2:ihacot} 
    
    \KwData{Set of pulses ($X$)} \medskip

    \KwFeatures{Frequency ($f_n$), Pulse width ($pw_n$), Level ($g_n$), Time of arrival ($t_n$)} \medskip

    \KwParameters{Minimum number of points to form a cluster ($\textrm{minPts}$), Confidence level ($\alpha$), Statistical test ($test$), Non-parametric method ($method$), Threshold ($\lambda$), distance ($distance$)} \medskip

     \KwProcedure{     
     \begin{enumerate}
     
        \item Pulses separation: apply HDBSCAN for all $x_i$ of $X$ from $t_n$, $f_n$, and $pw_n$ according to $\textrm{minPts}$: $C$ (Set of clusters) \medskip
        
        \item Pre-clusters aggregation: hierarchical agglomerative clustering based on $distance$ for all $c_i$ of $C$ from $\overline{f_n}$ and $\overline{pw_n}$: $P$ (Set of pre-aggregated clusters) \medskip     
        
       \item Cluster aggregation: apply hierarchical agglomerative clustering based on optimal transport distances presented in algorithm~\ref{algo_AP1:cluster_aggregation} considering $\alpha$, $test$, $method$ and $\lambda$ for all $p_i$ of $P$ from $t_n$ and $g_n$: $Y$ (Set of aggregated clusters) 
    
    \end{enumerate}}
    
    \KwResult{$Y$, Deinterleaving sets of pulses.} \medskip
       
\end{algorithm}


    

\medskip

\begin{figure}[h]
\centerline{\includegraphics[width = 9cm]{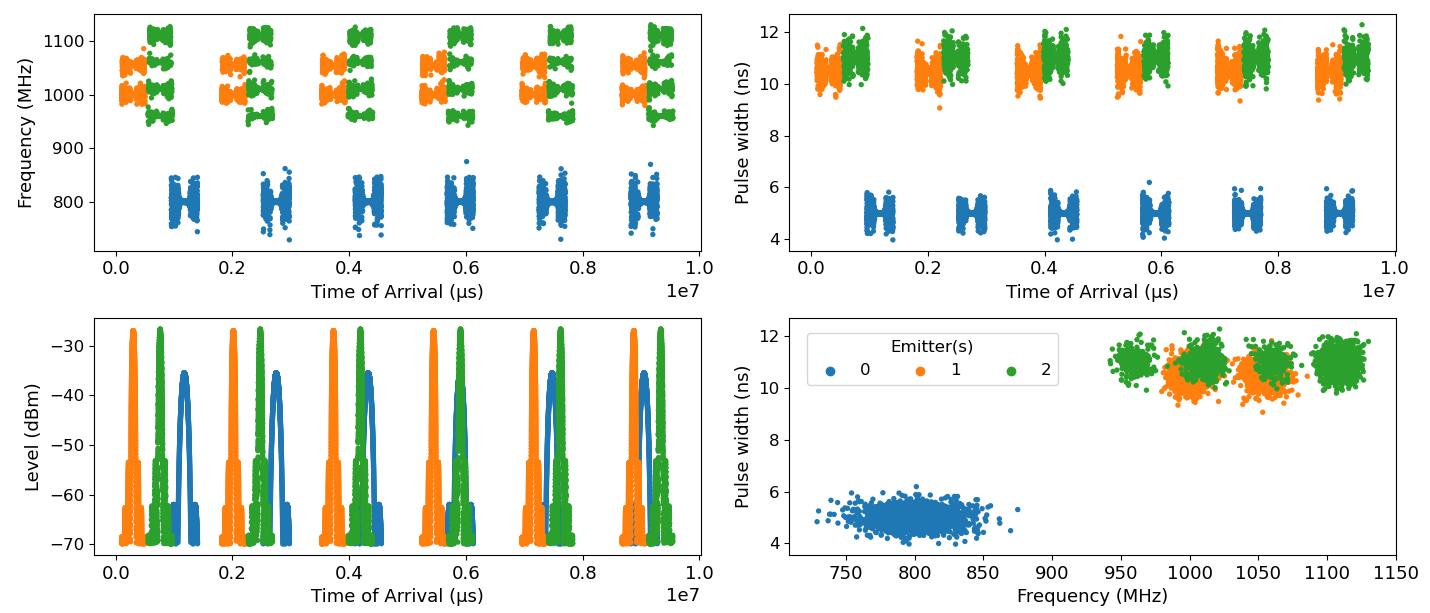}}
\caption{Example of a simulated signal gathering the pulses of three emitters identified by a color.}
\label{fig_AP2:step0_data_description}
\end{figure}

Fig.~\ref{fig_AP2:step0_data_description} shows an example of a simulated signal gathering pulses from 3 emitters. In the $(f_n, t_n)$ plane, the first emitter is easily identifiable and characterized by low frequencies (around 800 MHz). In comparison, the pulses of the two other emitters share the same frequency bands apart from 900 MHz. The pulse width spread and the simultaneity of emission of the two emitters emitting above 900 MHz, presented respectively in the $(pw_n, t_n)$ and $(g_n, t_n)$ planes, are not sufficient to differentiate them. The spreading of the pulses is caused by the presence of noise in the signal; this result is confirmed by looking at the pulses spread in the $(f_n, pw_n)$ plane with the pulses overlapping, leading to an inseparability of these two emitters.\\

\subsection{Grouping of pulses with the same characteristics in the $(t_n, f_n, pw_n)$ plane with HDBSCAN}\label{ssec:AP2step1}


{A three-dimensional vector $ X = \{(t_1, f_1, pw_1), (t_2, f_2, pw_2), ..., (t_N, f_N, pw_N)\}$ grouping time of arrival, frequency, and pulse width pulses from a signal of $N$ pulses is passed as input in HDBSCAN to be clustered.}

\begin{figure}[h]
\centerline{\includegraphics[width = 9cm]{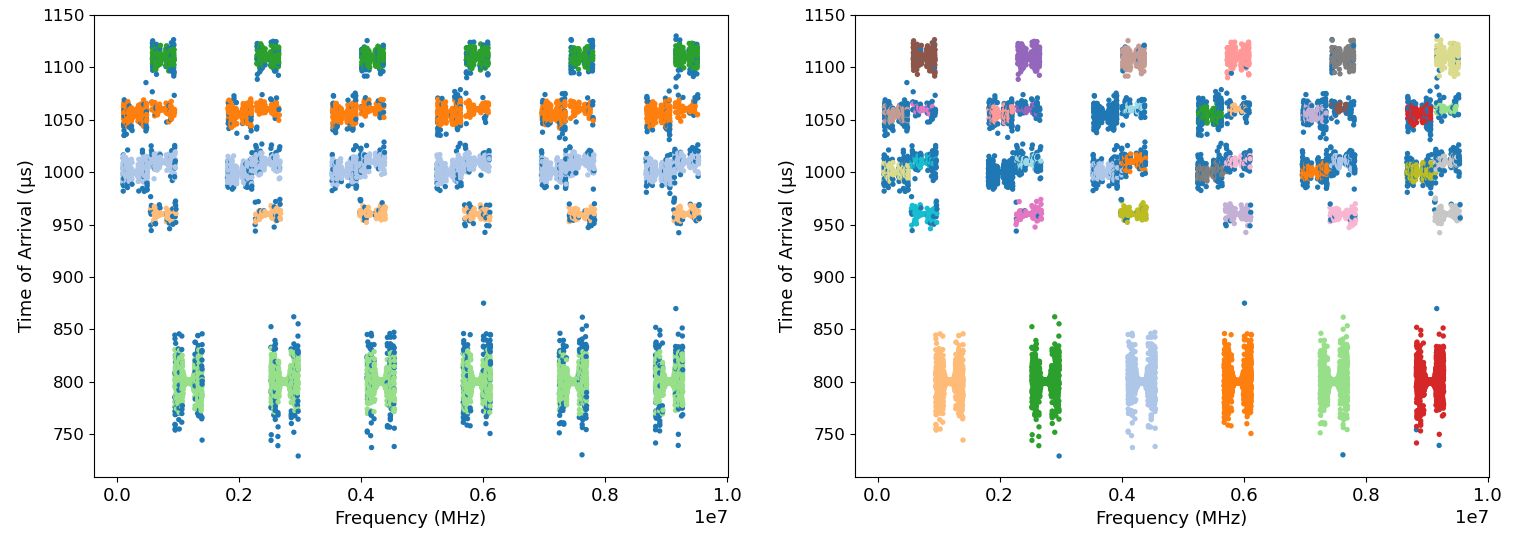}}
\caption{Comparison of HDBSCAN clustering respectively performed in 2 dimensions on the left plot and 3 dimensions on the right plot.}
\label{fig_AP2:step1_clustering_comparison}
\end{figure}

Incorporating the last feature in the clustering improves emitters' pulse discrimination, as shown in Fig.~\ref{fig_AP2:step1_clustering_comparison}. The clustering results obtained in 2 dimensions from $(f_n, pw_n)$ plane are displayed on the left plot; HDBSCAN identifies 5 clusters with an outliers class. The 2-dimensional clustering mixed the pulses of the two emitters at 1000 and 1060 MHz. Conversely, adding the third dimension improves the separation, as highlighted by the right plot; the algorithm identifies 42 clusters with a class of outliers (-1).

\subsection{Estimation of frequencies and pulse width means from previous clusters}\label{ssec:AP2step2}

\begin{figure}[h] 
    \centering \includegraphics[width = 9cm]{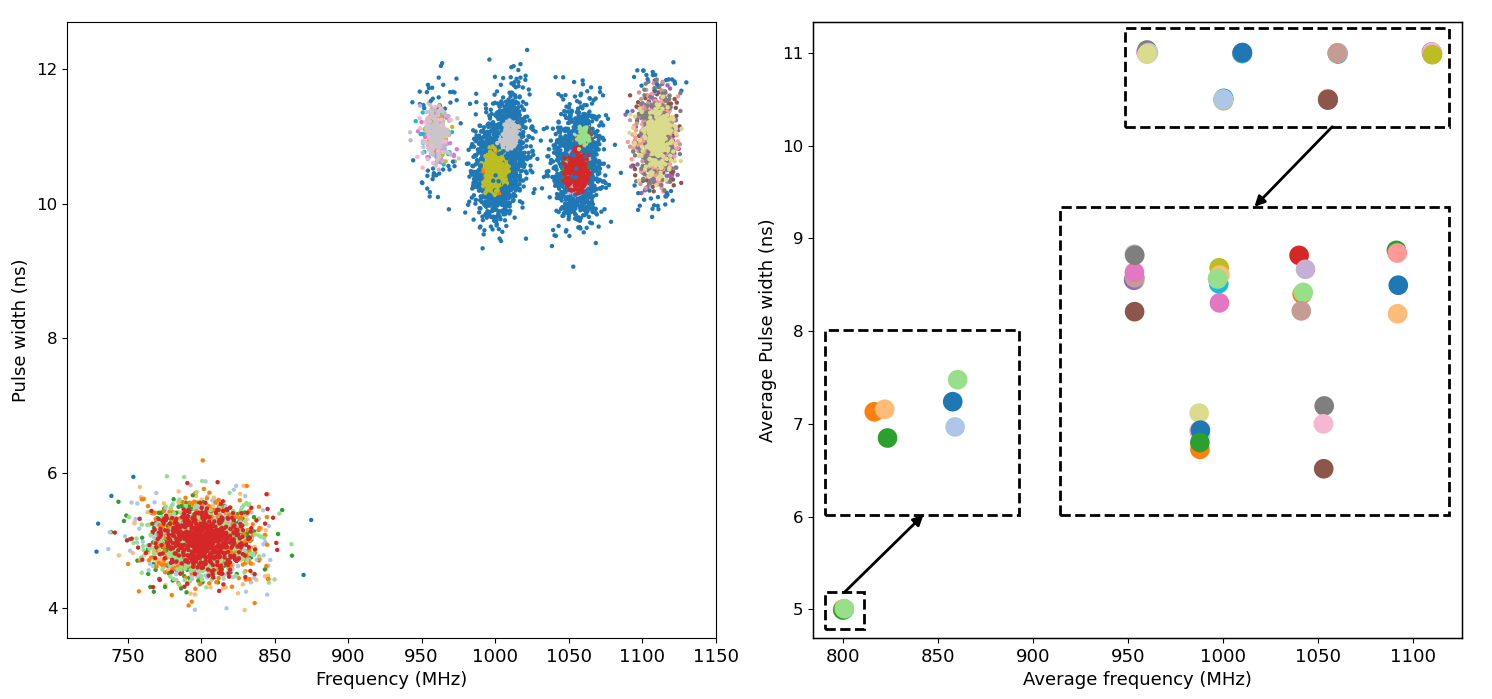}
\caption{HDBSCAN clustering results performed in 3 dimensions in the  $(f_n, pw_n)$ plane on the left and cluster averages in the $(f_n, pw_n)$ plane on the right. HDBSCAN detects 42 clusters with an outliers class (-1) identified by colors.}
\label{fig_AP2:step2_data_comparison}
\end{figure}

The clusters obtained in 3 dimensions by HDBSCAN displayed on the left graph of Fig.~\ref{fig_AP2:step2_data_comparison} are overlapped. We compute the previous clusters' frequency and pulse width averages to obtain a new representation of each cluster in the $(f_{mean}, pw_{mean})$ plane, as highlighted in the right plot. Each dot represents the clusters' frequency and pulse width averages. This new representation improves cluster separation and avoids overlapping cluster pulses. 


\subsection{Grouping clusters with the same characteristics from frequency.}\label{ssec:AP2step3}

At this step, each cluster is represented by a frequency and pulse width averages: $({f_{mean}, pw_{mean}})$. From this new representation, a classical hierarchical agglomerative clustering using Euclidean distances is applied to group clusters with the same characteristics. As previously, Algorithm~\ref{algo_AP1_algo:DecisonalModel_test} is applied, and fusions are stopped using the Kolmogorov-Smirnov test, checking that the frequency and pulse widths of the clusters follow the same distribution. Alternatively, the Student test\cite{student1908probable} can also be used. 



\begin{figure}[h]
\centerline{\includegraphics[height = 5cm]{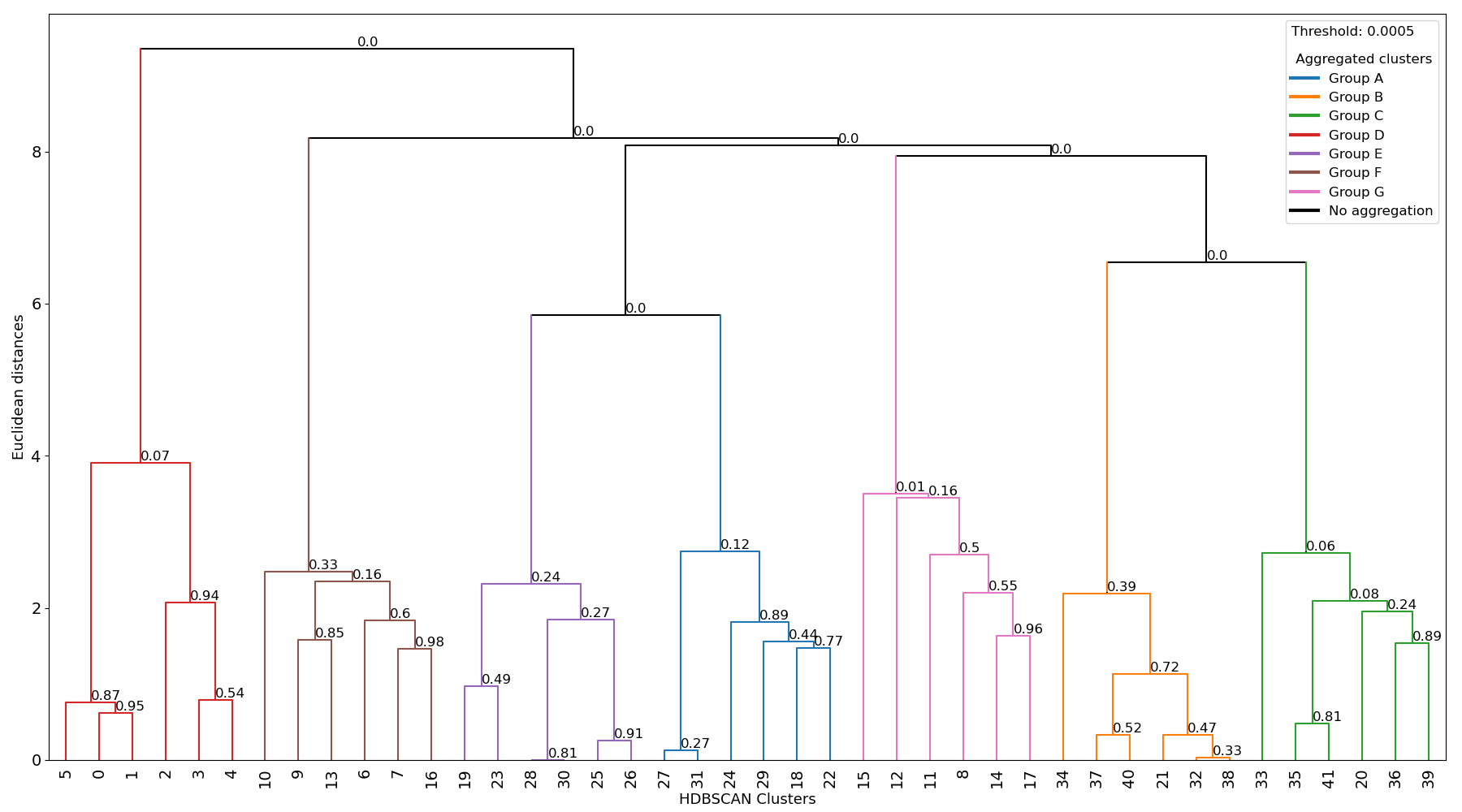}}
\caption{Dendrogram representing the aggregations of hierarchical agglomerative clustering. The $p$-values of the Kolmogorov-Smirnov test from $f_n$ evaluating the aggregated groups are displayed at each aggregation, while each color represents the aggregated cluster.}
\label{fig_AP2:step3_dendrogram}
\end{figure}

\begin{figure}[h]
\centerline{\includegraphics[width = 9cm]{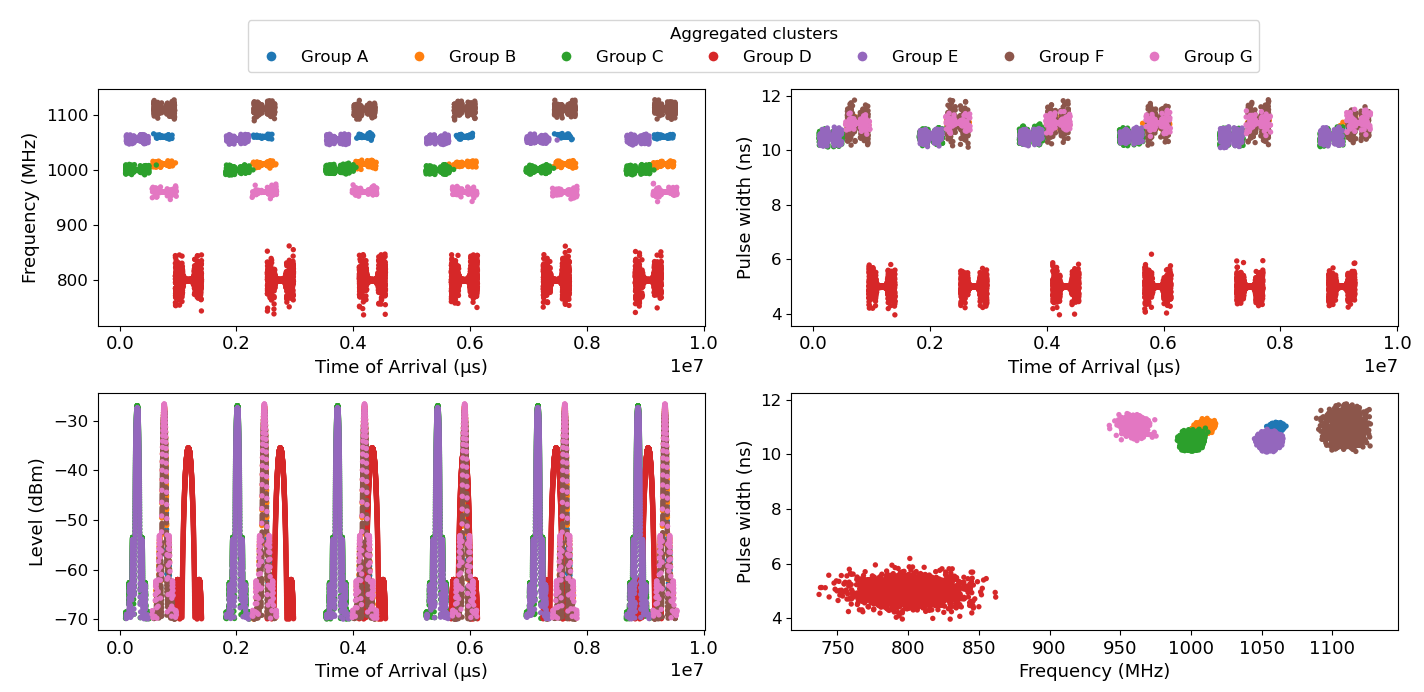}}
\caption{Aggregated cluster obtained following the dendrogram from hierarchical agglomerative clustering represented by a color.}
\label{fig_AP2:step3_data_aggregate_grouping}
\end{figure}

The dendrogram in Fig.~\ref{fig_AP2:step3_dendrogram} displays the groupings made from the pulses' characteristics by applying Algorithm~\ref{algo_AP1_algo:DecisonalModel_test} where the branches' heights are the optimal transport distance adjusted to improve the dendrogram. Here, the aggregated groups obtained using the Student test coincide with those obtained with the Kolmogorov-Smirnov test. The decision model identified seven aggregated clusters, and the results are presented in Fig.~\ref{fig_AP2:step3_data_aggregate_grouping}.

\subsection{Clusters agglomeration with similar temporal characteristics with optimal transport distances}\label{ssec:AP2step4}

\begin{figure}[h]
\centerline{\includegraphics[width = 9cm]{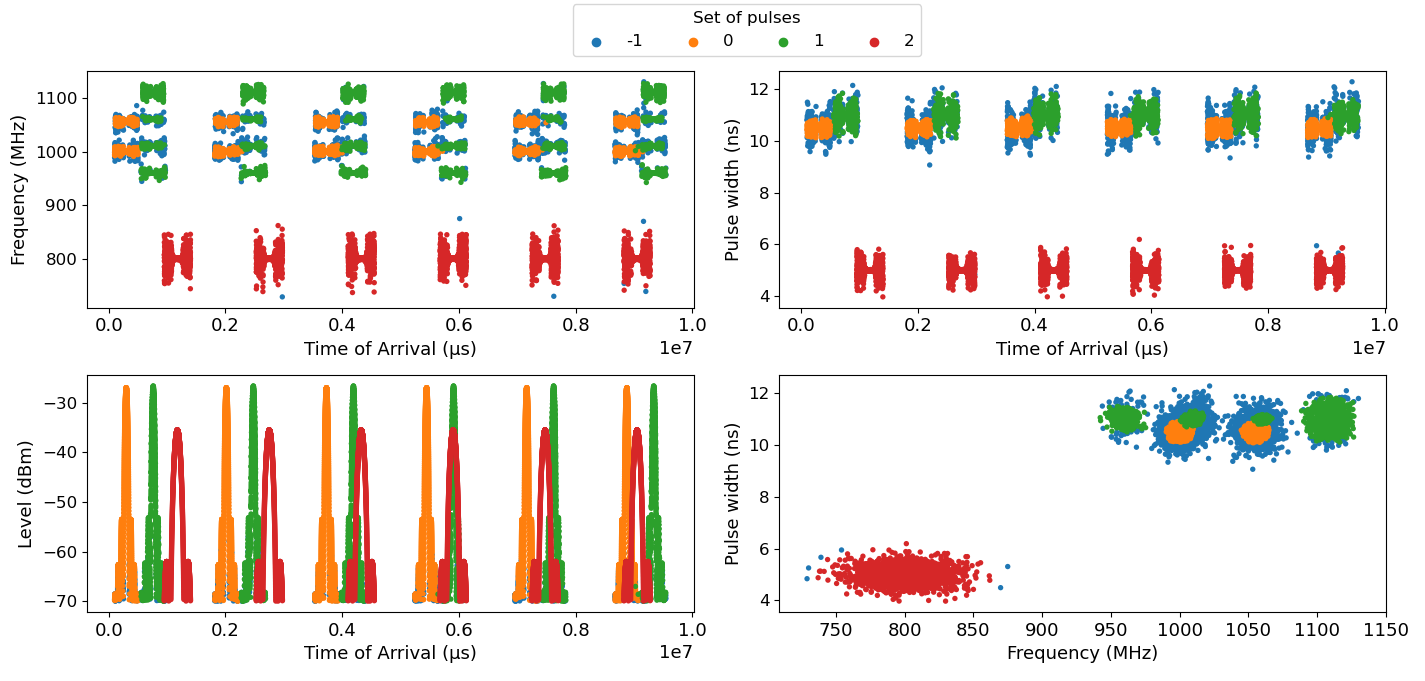}}
\caption{Final grouping after applied hierarchical agglomerative clustering combined with optimal transport distances and kernel density estimation on excluded clusters. Each color identified a set of pulses with an outliers class (-1).}
\label{fig_AP2:step4_data_final}
\end{figure}

The last step of this algorithm is similar to the previous method, based on hierarchical clustering with optimal transport distances. Fig.~\ref{fig_AP2:step4_data_final} shows the final three sets of pulses that our method identified and a class of outliers (-1). The methodology could separate and regroup emitters' pulses with similar characteristics in the presence of noise; HDBSCAN excluded about 9\% of pulses.


\subsection{{Computational complexity}}\label{ssec:AP2step5}

{The complexity of this second method is the sum of the complexity of HDBSCAN, the pre-cluster aggregation, and the cluster aggregation. The complexity of HDBSCAN and the cluster aggregation is identical to the first method, with $C$ now being the number of pre-aggregation clusters. The pre-cluster aggregation is a simple hierarchical clustering, with complexity $O(P^3)$, with $P$ the number of clusters identified by HDBSCAN. $P$ can be estimated by $P = CTS$, where $S$ is the average number of sweep by second. Compared to the previous algorithm, the complexity is augmented by $O(C^3T^3S^3)$, or $O( \bar C^3 K^3 T^3 S^3)$.

}

%% file: part_results.tex
\section{Results}\label{ssec:results}

\begin{figure}[h]
\centerline{\includegraphics[width = 9cm]{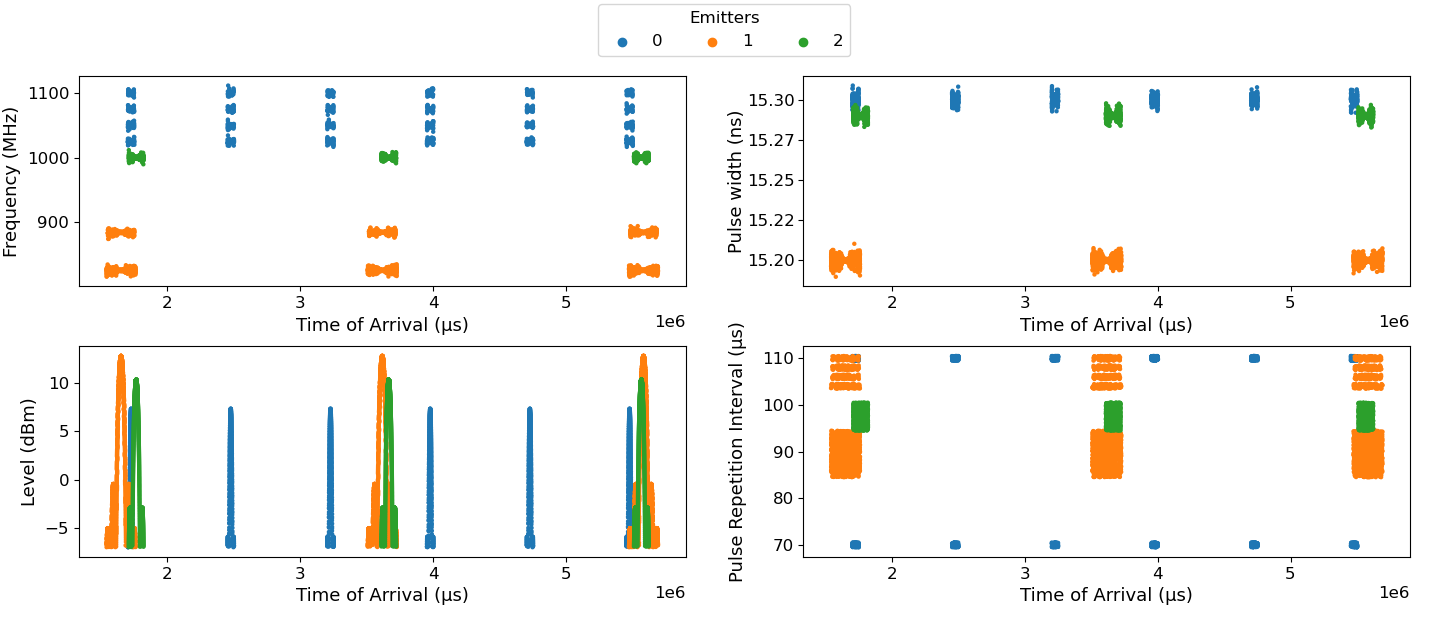}}
\caption{Simulated signal gathering 10044 pulses from 3 emitters. Each color identified an emitter.}
\label{fig_Results:data_simple}
\end{figure}

The two proposed methods are tested on a simulated signal with three emitters, with characteristics given in tab~\ref{tab:characteristics}. An example with 10044 pulses is shown in Fig.~\ref{fig_Results:data_simple}. The characteristics of the emitters, with multiple frequencies, similar pulse widths, and PRIs, are chosen to highlight the robustness of the methods concerning adverse cases. The methods' performance is assessed using the Adjusted Rand Index (ARI)~\cite{hubert1985comparing}. As the emitters can emit on several frequencies with various pulse widths, methods based on clustering on these features will fail to group all pulses of a given emitter.

\subsection{Comparison with PRI-based methods}

\begin{figure}[h]
\centerline{\includegraphics[width = 9cm]{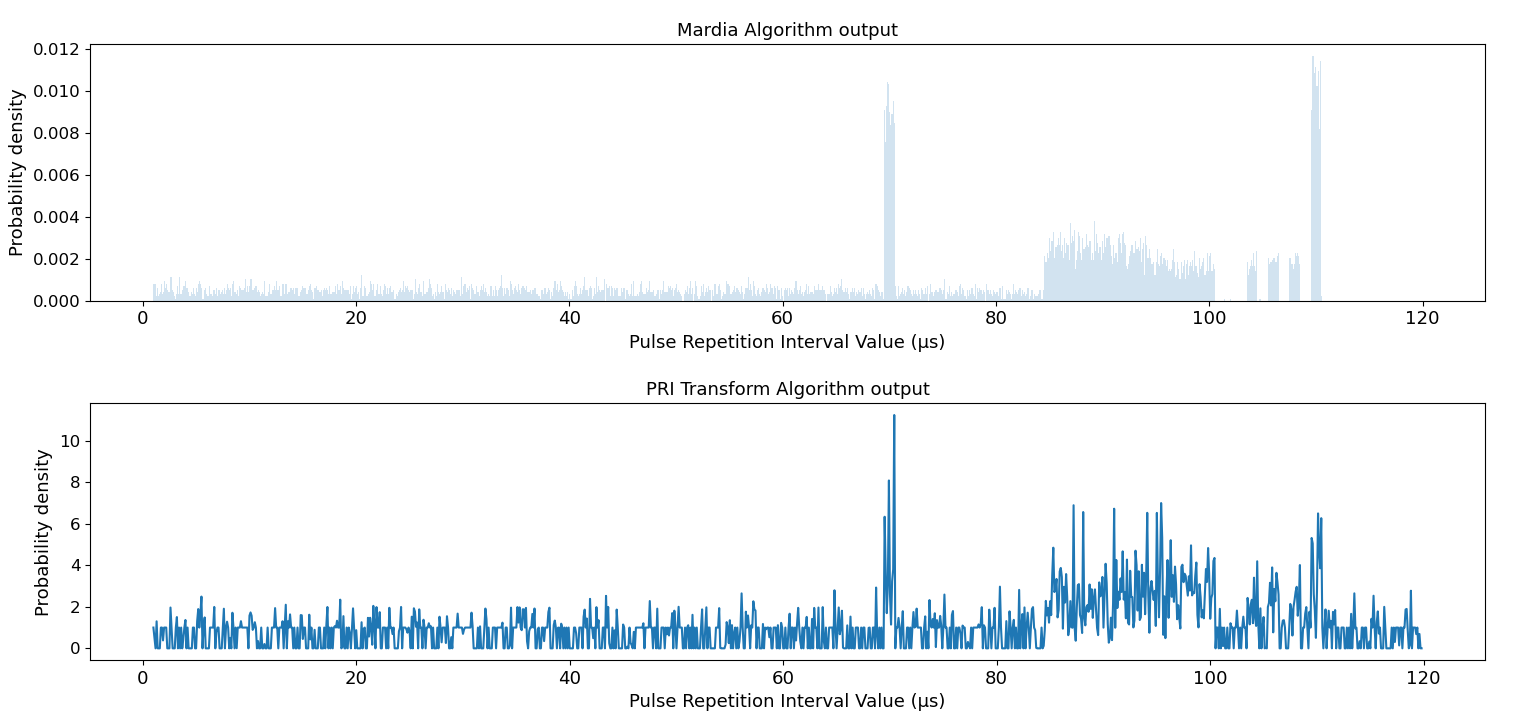}}
\caption{Results of Mardia and PRI Transform algorithms to deinterleave the signal of the Fig.~\ref{fig_Results:PRI_based_methods} from the PRI.}
\label{fig_Results:PRI_based_methods}
\end{figure}

Two approaches were selected to be compared to the two deinterleaving methods developed. One conventional method, based on the analysis of PRI histograms~\cite{mardia1989new} and another, more recent, based on the PRI transform~\cite{nishiguchi1983new}, have been selected. Methods based on Deep Learning models have been deliberately set aside as they require a lot of data for the training step and perform inadequately with few data, while the proposed method is unsupervised. The outputs of these methods, shown in Fig.~\ref{fig_Results:PRI_based_methods}, cannot be used to estimate the number of emitters and separate their pulses. Indeed, the complex PRI patterns of each emitter and the similarity of their PRIs are such that both methods fail.

\begin{table}[h]
\centering
\begin{tabular}{|c|c|c|c|}
\hline
Emitter & Frequency (MHz) & Pulse width (ns) & PRI ($\mu$s) \\ \hline
0 & \begin{tabular}[c]{@{}c@{}}1025, 1050, \\ 1075, 1100\end{tabular} & 15.3 & 70, 110 \\ \hline
1 & 825, 884 & 15.2 & \begin{tabular}[c]{@{}c@{}}85, 86, 87, 88, \\ 89, 9, 91, 92,  \\ 93,  94, 104, 106, \\ 108, 11.\end{tabular} \\ \hline
2 & 860 & 15.3 & \begin{tabular}[c]{@{}c@{}}95, 96, 97,  \\ 98, 99, 100.\end{tabular} \\ \hline
Std & 2.14 & 1e-3 & 2.55\\ \hline
\end{tabular}
\caption{Simulated emitters characteristics.}
\label{tab:characteristics}
\end{table}


Fig.~\ref{fig_Results:PRI_based_methods} represents the result of Mardia's and PRI Transform method applied to our previous signal. The calculated PRI on Mardia's method values visually oscillates around 70$\mu $s. We also observe a block of PRI between 80 and 110 $\mu $s. The same results are observable in the PRI Transform method but are less pronounced. The application of the PRI Transform method is significantly time-consuming due to the length of the signal; we restricted the search for PRI values to a predefined range instead of considering all time of arrival values. Despite knowing the PRI values to detect, we failed to find the emitter's emission patterns. As emitters share emission patterns, using these methods to estimate the present PRI values and detect the emitters in this signal is unfeasible.\\


\begin{figure}[htbp]
\centerline{\includegraphics[width = 9cm]{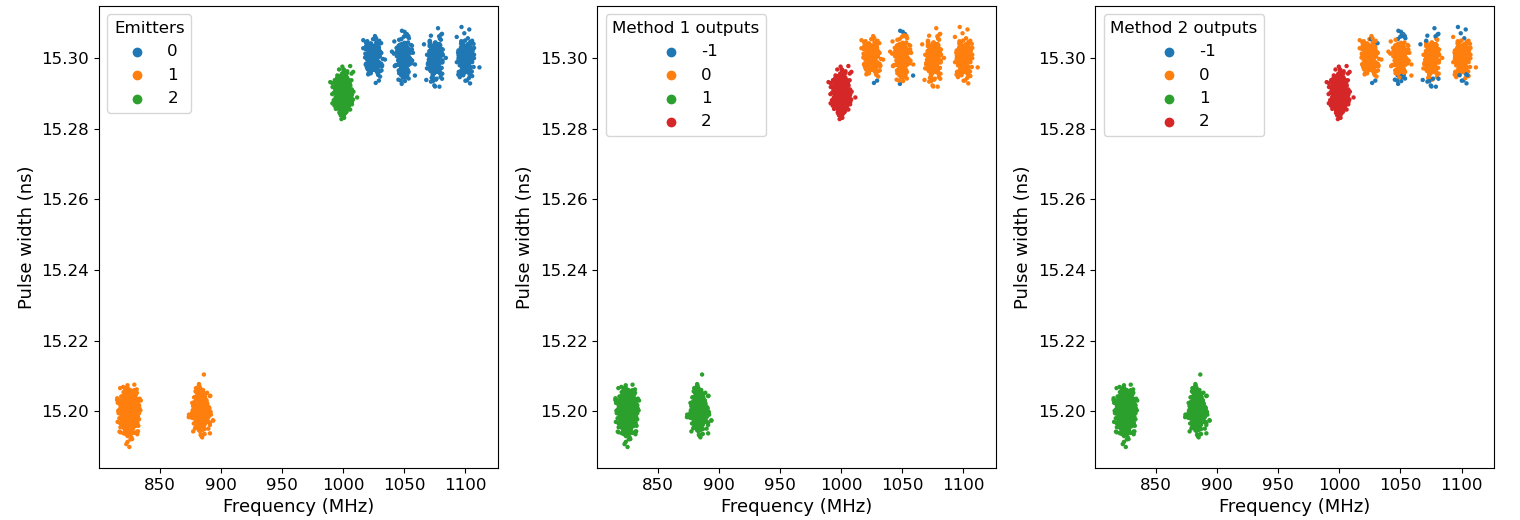}}
\caption{Results of the HACOT and IHACOT developed in sections~\ref{ssec:AP1step0} and \ref{ssec:AP2step0} to deinterleave the signal of the Fig.~\ref{fig_Results:data_simple} based on Kolmogorov-Smirnov test for all grouping phases.}
\label{fig_Results:Mtr_comparison_freqxpw}
\end{figure}

To facilitate understanding, HACOT and IHACOT identify the deinterleaving methods based on the 2D and 3D clustering developed and presented in sections~\ref{ssec:AP1step0} and ~\ref{ssec:AP2step0}. The methods are performed based on applying the Kolmogorov-Smirnov test on all grouping phases, and results are displayed in Fig.~\ref{fig_Results:Mtr_comparison_freqxpw}. The emitters have different characteristics, and the pulses are sufficiently spread in the plane, leading to a fast and correct deinterleaving of the three emitters' pulses. HACOT detects 0.01\% outliers against 0.5\% for IHACOT.\\

\subsection{Robustness to outliers}

\begin{figure}[h]
\centerline{\includegraphics[width = 9cm]{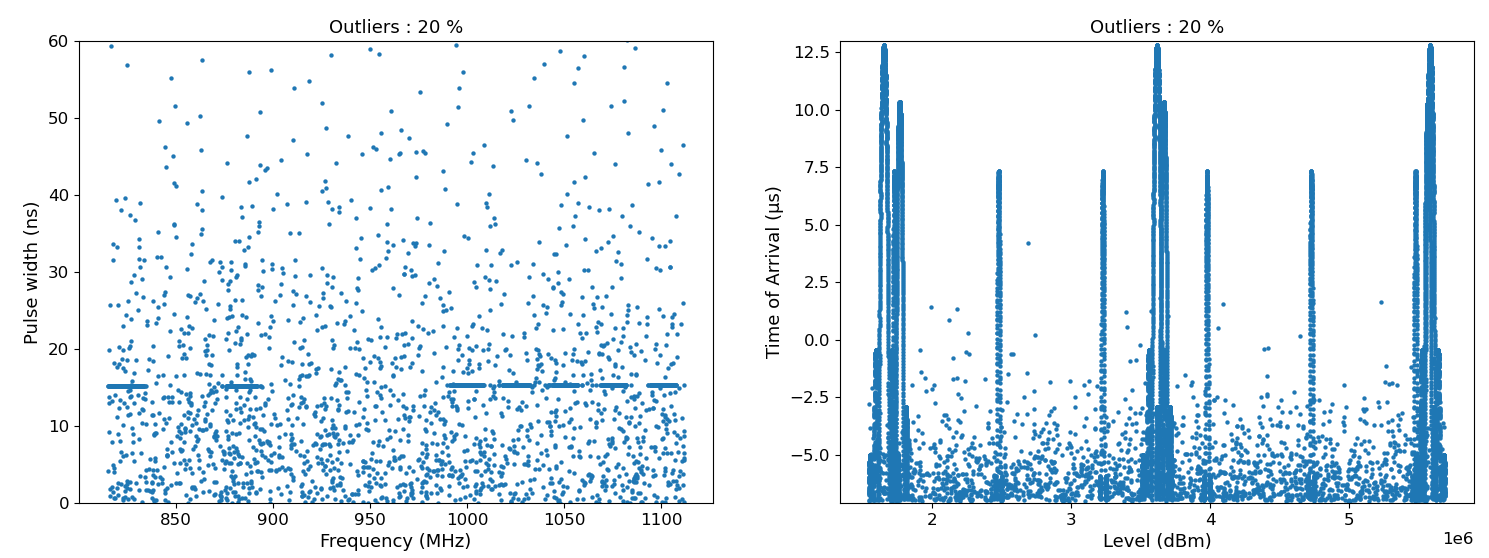}}
\caption{Pulses representation in the $(f_n, pw_n)$ and $(t_n, g_n)$ planes according 20\% of outliers added in the initial signal.}
\label{fig_Results:Outliers_plots}
\end{figure}

The performance of the methods with varying proportions of outliers is evaluated. The proportion of outliers added to the signal varies between 0 and 90\%. The performance criteria are averaged over 50 realizations. The outliers were generated uniformly between the range values for frequency and time of arrival. Alternatively, an exponential law generates pulse widths and level values because it favors the appearance of low values, making the outliers' values more realistic, as illustrated in Fig.~\ref{fig_Results:Outliers_plots}, which represents the previous signal in the $(f_n, pw_n)$ and $(g_n, t_n)$ planes with 20\% of outliers added.\\

\begin{figure}[h]
\centerline{\includegraphics[width = 9cm]{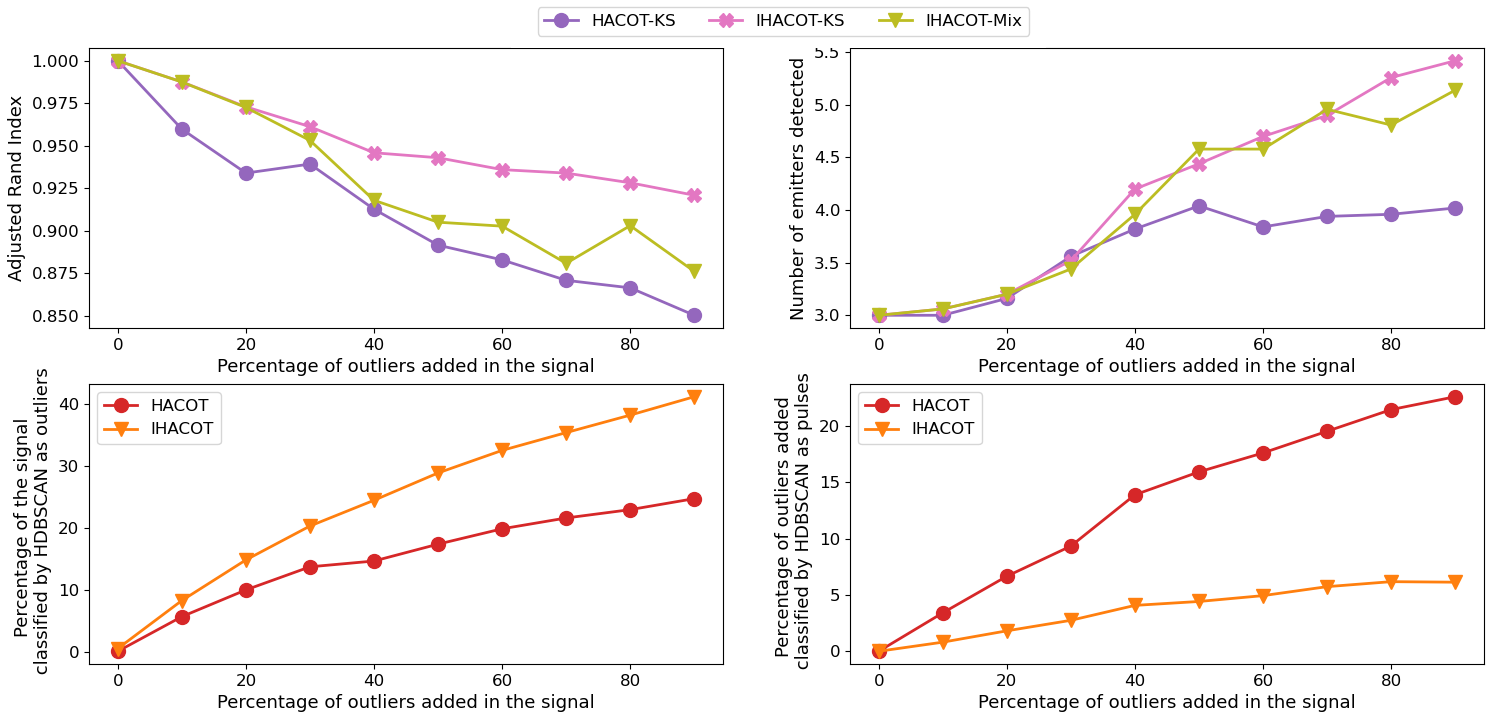}}
\caption{Performance of deinterleaving methods according to the outliers rate added with ARI, number of emitters detected, part of the signal classified by HDBSCAN as outliers, and part of outliers added in the signal classified in the sets of pulses. Each curve identifies a method.}
\label{fig_Results:Outliers_metrics}
\end{figure}

The previously mentioned metrics used to evaluate the algorithm's performances are presented in Fig.~\ref{fig_Results:Outliers_metrics}. IHACOT is evaluated through two approaches: using only the Kolomogorov-Smirnov test for the cluster regrouping phases (IHACOT-KS) and by mixing the Student test for the first grouping and the Kolmogorov-Smirnov test for the second grouping (IHACOT-Mix). IHACOT-KS and ICAOT-Mix give higher results, as the ARI indicates. The better performance of these methods is due to the 3D-clustering, which cleans the data more efficiently by detecting outliers, leading to more reliable and stable results. The result is confirmed by the graph below: the number of pulses classified as outliers by HDBSCAN is higher for IHACOT-KS and IHACOT-Mix than HACOT-KS as the outliers added in the signal increase. Moreover, the final sets of pulses obtained from IHACOT-KS and IHACOT-Mix are composed of fewer outliers than those from HACOT-KS, as illustrated by the bottom right plot, meaning the 3-dimensional clustering provides less noisy sets of pulses. IHACOT-KS and IHACOT-MIX overestimate the number of emitters detected in the signal, as shown in the top-right plot. For our problem, we prioritized the overestimated emitter detected rather than underestimating them and having mixed pulses in a single cluster. IHACOT-KS and IHACOT-Mix seem more adapted to make deinterleaving.\\




\begin{figure}[h]
\centerline{\includegraphics[width = 9cm]{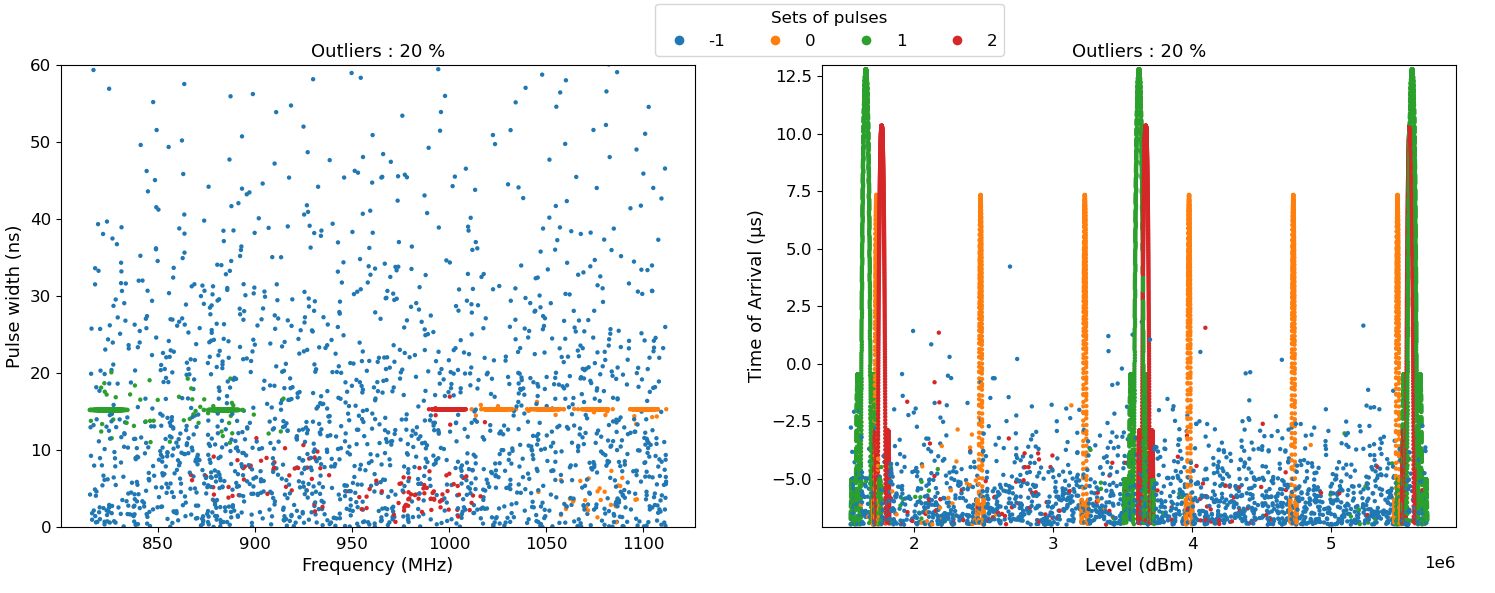}}
\caption{Results of IHACOT with Kolmogorov-Smirnov for each grouping phase applied with 20\% of outliers added in the $(f_n, pw_n)$ and $(t_n, g_n)$ plane. Each color represents a detected emitter, and -1 is the outliers class.}
\label{fig_Results:Outliers_deinterleaving}
\end{figure}

\medskip

Fig~\ref{fig_Results:Outliers_deinterleaving} show the results of IHACOT-KS applied on Fig.~\ref{fig_Results:Outliers_plots}. Note that some outliers corresponding to low-intensity pulses are grouped in a set of pulses. By employing 3D clustering, the pulses emitted were successfully separated, avoiding mixing pulses from multiple emitters into one group despite adding outliers in the signal.

\subsection{Sensitivity to estimation errors}

Similarly, we applied the same process to evaluate the performance of our algorithms by varying the noise level. The noise level in the estimated parameters of the PDW is varied by multiplying the baseline level given in Table~\ref{tab:characteristics} with a factor(\textit{noise coefficient}) ranging from 1 to 10. Likewise, 50 simulated signals per noise level were generated in the outliers experimentation to average the results and obtain consistent results.\\



\begin{figure}[h]
\centerline{\includegraphics[width = 9cm]{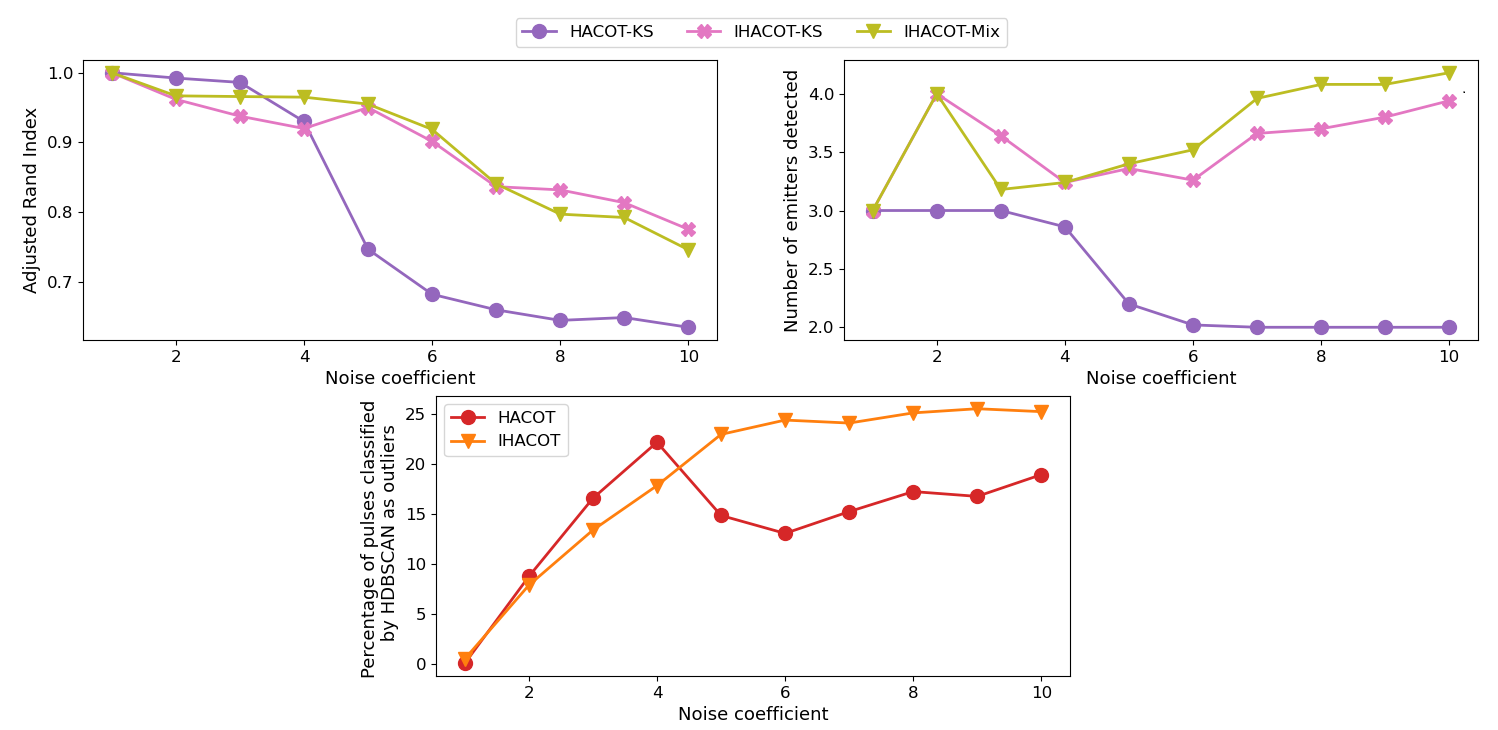}}
\caption{Performance of deinterleaving methods according to the noise added with ARI, number of emitters detected, and part of pulses classified by HDBSCAN as outliers. Each curve identifies a method.}
\label{fig_Results:Noise_metrics}
\end{figure}

Fig.~\ref{fig_Results:Noise_metrics} shows the evolution of previously used metrics, averaged over 50 realizations, applied on Fig.~\ref{fig_Results:data_simple} with different noise levels. ARI is high for all methods, and their results are similar. IHACOT-KS and IHACOT-Mix give better results than HACOT-KS. Methods based on 3D clustering classify more pulses as outliers, leading to better separation and avoiding mixing pulses in the numerous 3D detected clusters, as shown at the bottom. Finally, as illustrated in the top-right plot, HACOT-KS mixes the emitter's pulses given a certain noise level, while IHACOT-KS and IHACOT-Mix tend to overestimate them slightly.\\

\begin{figure}[h]
\centering
\includegraphics[width = 9cm]{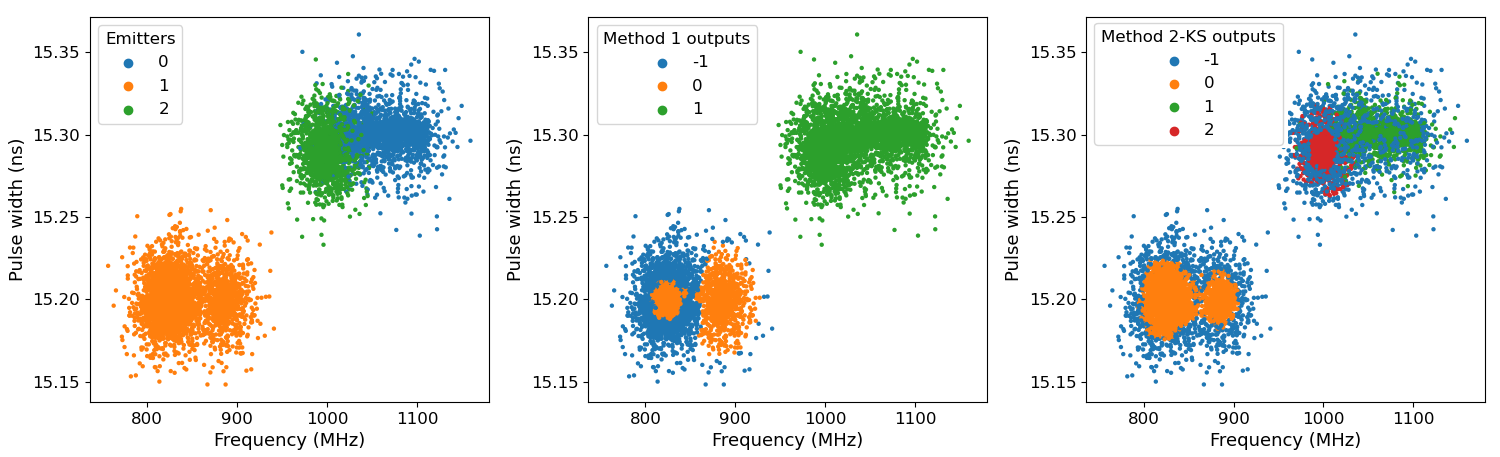}
\caption{Deinterleaving results in the plane $(f_n ,pw_n)$ obtained from HACOT-KS and IHACOT-KS applied on figure~\ref{fig_Results:data_simple} with the baseline level multiplied by the coefficient 40.}
\label{fig_Results:noise_deinterleaving}
\end{figure}


An application of HACOT-KS and IHACOT-KS is shown in Fig.~\ref{fig_Results:noise_deinterleaving}. The left plot represents the ground truth, and the middle and the right plots are the results of HACOT-KS and IHACOT-KS. Each color represents an emitter and (-1) the outliers. IHACOT-KS separates the pulses better by applying a 3D clustering, whereas HACOT-KS fails to separate the pulses of the two emitters around 15.3 ns. 

%% file: part_conclusion.tex
\section{Conclusion}


This paper proposes two deinterleaving methods capable of managing complex emitters without considering the PRI pattern. Hierarchical agglomerative clustering combined with optimal transport distances is used to identify emitters using several frequencies, presented agility characteristics, or having multiple operation modes. Here, the assumption is that the pulses of a given emitter, with varied parameters, are received at similar times. The PRI is not used, making the methods robust with respect to complex PRI patterns. However, using frequency and pulse width in the first clustering step is straightforward, specifically when the listening perimeter concerns a port or an airport, and the signal is noisy; a second approach, including pre-clustering made from the time of arrival, frequency, and pulse width, is initiated, to separate pulses better, before applying the previous hierarchical agglomerative clustering using optimal transport distances developed to overcome these limitations. The two resulting algorithms are flexible, playing with various hyperparameters (\textit{e.g.}, threshold, kernels, etc.) that are adjusted with the datasets.  Finally, simulation results have shown the interest of the proposed methods in terms of accuracy. Furthermore, in addition to the robustness concerning complex emission patterns, the proposed approaches are also robust with respect to the estimation of the parameters and outliers.\\

The deinterleaving approaches are based on accessible and reliable features. The deinterleaving results could be improved by including additional features, such as the direction of arrival, which is not always accessible but highly discriminating. A complementary algorithm could be developed based on temporarily available features and combined with the principal methodology to improve the deinterleaving of a signal. The Kolmogorov-Smirnov and Student tests only apply between two data distributions during the hierarchical approach to prune the dendrogram, implying that at least two emitters are present in the deinterleaving signal. Other tests or methods could be explored to consider cases where the intercepted pulses correspond to only one emitter or combine other relevant tests to improve the pruning. {The similarity between emitters strongly influences the approaches, specifically when the emitters share common characteristics. In the electronic warfare context, operators deliberately select listening areas; specific areas like airports or ports are not typically favored for these activities, although these locations can potentially pick up many emitters' signals. Our considerations have not accounted for scenarios where a hundred or more emitters might be within the surveillance area but could be subject to extending this research. During the hierarchical approach, the threshold evaluating the significance of clusters is deliberately set at 100, considering the results we have seen on several signals. We are currently working on the development of a statistical method to set this threshold more confidently. The main idea will be to compare the distribution of the cluster, grouping all its pulses with distributions by removing pulses first using the Kolmogorov-Smirnov test to determine if the original distribution is statistically different from the distribution with suppressed pulses to fix the threshold and several descriptive statistics (comparison of means, standard deviations, kurtosis, skewness, etc.).} Finally, applying the hierarchical clustering algorithm using the optimal transport distances assumes that the clusters belonging to an emitter are simultaneously active, which in some instances is not valid.

